\documentstyle[12pt]{article}
\renewcommand{\theequation}{\thesubsection.\arabic{equation}}
\begin{document}
\pagestyle{plain}
\pagenumbering{arabic}
\baselineskip=20pt

\title{AN ALTERNATIVE APPROACH FOR THE DYNAMICS OF POLARONS IN ONE DIMENSION}

\vspace*{1.0cm}

\author{A.H. Castro Neto
\and A.O. Caldeira\\
Instituto de F\'{\i}sica ``Gleb Wataghin"\\
Departamento de F\'{\i}sica do Estado S\'{o}lido e Ci\^{e}ncia dos Materiais\\
Universidade Estadual de Campinas\\
13081 Campinas, SP, Brazil}

\date{14 August 1991}

\maketitle

\setcounter{page}{1}
\begin{abstract}

\vspace*{2.0cm}

\indent
We developed a new method based on functional integration to treat
the dynamics of polarons in one-dimensional systems. We treat the
acoustical and the optical case in an unified manner, showing their
differences and similarities. The mobility and the diffusion coefficients
are calculated in the Markovian approximation in the strong coupling limit.

\end{abstract}

\newpage

\section{Introduction}

\indent
The main task of this paper is to develop a new formalism to treat
the dynamics of acoustical or optical polarons in the strong
coupling limit. Although we are applying the formalism to a specific
problem we think that it can also be applied to a large class of
phenomena, specially problems involving quantization of zero
frequency modes in theories which have solitons as solutions of their
semiclassical equations of motion.

\indent
When we have a particle (electron) interacting with a given background
(phonons in this case) and wish to study its effective dynamics, it is
now well-known that we must trace over the phonon coordinates and study
the time evolution of the reduced density operator of the electronic
system. However, in the specific case of electron-phonon coupling, the
``effective propagator" for this operator is extremely cumbersome, preventing
us to get any simple result out of this standard analysis. Therefore, one
should search for an extra step before blindly tracing the phonon coordinates
out of the problem.

\indent
We start from a very intuitive point, treating the electron-phonon
Hamiltonian as a ``semiclassical" Hamiltonian \cite{dav}.
This ``semiclassical" picture provides us with solutions which are solitons,
that is, solutions which do not change their shape with time. These solitons
will be the basic entities for the future solution of the problem. We show that
   the best
basis in which one can expand the field operators of the quantum Hamiltonian is
obtained from the problem of an electron trapped in a self consistent
potential well. We will call it the ``adiabatic basis" since the
strong coupling limit is the adiabatic limit (see section 2).

\indent
Once we have obtained the Hamiltonian in the adiabatic form we can
eliminate the electronic part perturbatively, that is, we trace
over the electron coordinates. This treatment gives
rise to an effective Hamiltonian for the phonon system which has
renormalized phonons and a zero frequency mode, namely, the polaron.
An important feature of this Hamiltonian is that it can be straightforwardly
generalized for a non-interacting many electron system .

\indent
Using the well-known ``collective coordinate formalism" we transform
the effective Hamiltonian into a Hamiltonian of a particle, the
polaron, coupled to a new set of phonons.
It can be shown that for a small polaron momentum the problem can be
put in a very simple form.

\indent
Actually, this is a more systematic way to apply the ideas used by Schuttler
and Holstein \cite{schu} to a quantum dissipation problem as the necessary
step before the tracing procedure. Here, we shall repeat part of their
arguments for the sake of completeness.

\indent
 Finally, using the functional integral formalism, we can show that
the polaron behaves as a brownian particle due to the scattering of
phonons. So, we have developed a method which allows us to calculate the
physical quantities of interest, such as the damping parameter (mobility)
and the diffusion coefficient without appealling to kinetic theory.

\indent
In section II we will present the model and exhibit the adiabatic basis
for the strong coupling limit while in section III we obtain the effective
Hamiltonian for the polaron coupled to the renormalized phonons.
In section IV we use the functional integral formalism in order to
show how this problem can be treated as a brownian motion problem and in
section V we use the previous results to calculate
the physical quantities of interest.
Section VI contains our conclusions.

\section{The Polaron Model and the Adiabatic Expansion}

\indent
Since in this paper we will treat the problem of an electron coupled to
acoustical or optical lattice vibrations (phonons) in an unified manner,
we decided to develop these two problems in parallel, in order to
show their differences and similarities.

\subsection{Optical Case}

\setcounter{equation}{0}

\indent
The optical polaron model is based on the Fr\"{o}hlich
Hamiltonian \cite{froh} for electrons coupled to longitudinal optical
phonons. This
Hamiltonian can be written in the second quantized form, in one
dimension, as

\begin{equation}
H_o = \int{dx} \left\{ \frac{\hat{\pi}^2}{2\nu}+\frac{\nu\omega^2_o}{2}
\hat{\eta}^2 + \frac{\hbar^2}{2m}
\frac{\partial\hat{\psi}^{\dag}}{\partial
x}\frac{\partial\hat{\psi}}{\partial x} + \frac{D}{a} \hat{\eta} \:
\hat{\psi}^{\dag} \hat{\psi} \right\}
\end{equation}

\noindent
here $\hat{\pi}$ and $\hat{\eta}$ are the momentum and position
operator for the phonon field and $\hat{\psi}^{\dag}$ and $\hat{\psi}$ the
creation and destruction operators, respectively, for the electron
field. They obey the following commutation rules

\[
[\hat{\eta}(x,t),\hat{\pi}(x',t)] = i\hbar \delta(x-x')
\]
\[
\left\{\hat{\psi}(x,t), \hat{\psi}^{\dag}(x',t)\right\} = \delta(x-x')
\]

\noindent
where [ , ] denotes commutation and $\{ ,\}$ anticommutation. All the
other commutation (or anticommutation) relations are zero.

\indent
In (2.1.1), $\omega_o$ is the frequency of the phonons, $\nu = M/a$ is the
lattice density, $M$ the ion mass, $a$ the lattice parameter,
$m$ is the effective mass for the electrons in the conduction
 band and $D$ the coupling constant.

\indent
In order to analyze the physical content of (2.1.1) we will treat the
operators as ordinary functions and interpret the electron field as
the wave function for one electron in the lattice. In other words we
would say that we are treating the problem in the mean field approximation
where the operators are replaced by their mean values over configurations.
It would be emphasized that this is not an exact calculation but just an
artifact to obtain the best basis in which we would expand the operators
of the Hamiltonian (2.1.1) in order to get the strong coupling regime.
It can be easily
shown that the following Lagrangean can generate the Hamiltonian (2.1.1)

\[
L = \int{dx} \{  \frac{\nu}{2} \left( \frac{\partial \eta}{\partial
 t} \right)^2 -
\frac{\nu \omega^2_o}{2} \eta^2 + i\hbar \left( \psi^* \frac{\partial
\psi}{\partial t} - \psi \frac{\partial \psi^*}{\partial t} \right)
\]
\begin{equation}
- \frac{\hbar^2}{2m} \frac{\partial \psi^*}{\partial x} \frac{\partial
\psi}{\partial x} - \frac{D}{a} \eta \psi^* \psi \}
\end{equation}

\noindent
where $\psi$ is normalized:

\begin{equation}
\int{dx} \mid \psi(x,t) \mid^2 = 1 \, \, .
\end{equation}

\indent
The equations of motion for the Lagrangean (2.1.2) are

\begin{equation}
i\hbar \frac{\partial \psi}{\partial t} + \frac{\hbar^2}{2m}
\frac{\partial^2 \psi}{\partial x^2} - \frac{D}{a} \eta \psi = 0
\end{equation}
\begin{equation}
\frac{\partial^2 \eta}{\partial t^2} + \omega^2_o \eta + \frac{D}{M}
\mid \psi \mid^2 = 0
\end{equation}

\noindent
Equation (2.1.4) is the Schr\"{o}dinger equation for an electron in a
potential given by

\[
V(x,t) = \frac{D}{a} \eta(x,t)
\]

\noindent
while equation (2.1.5) is an equation for an oscillator with frequency
$\omega_o$ forced by the presence of an external field, $\mid \psi
\mid^2$. The picture is that of an electron which distorts the lattice
which, in its turn, produces a potential for the electron; a self
consistent interaction.

\indent
We are interested only in stationary solutions for the electrons,
that is, solutions of the form

\begin{equation}
\psi(x,t) \equiv \phi_o(x) e^{-iE_ot/\hbar}
\end{equation}

\noindent
as well as in static solutions for the lattice (adiabatic solution)

\begin{equation}
\eta(x,t) \equiv \eta_o(x)
\end{equation}

\noindent
 From (2.1.4), (2.1.5), (2.1.6) and (2.1.7) we get

\begin{equation}
\left(-\frac{\hbar^2}{2m} \frac{d^2}{dx^2} + \frac{D}{a} \eta_o(x)\right)
\phi_o(x) = E_o \phi_o(x)
\end{equation}
\begin{equation}
\eta_o(x) = - \frac{D}{M\omega^2_o} \phi^2_o(x)
\end{equation}

\indent
We can think of (2.1.8) and (2.1.9) as follows: we put the electron in
the lattice and the latter adjusts itself to the presence of the
former. As consequence the electron is trapped by the potential
well formed around itself.

\indent
Substituting (2.1.9) in (2.1.8) we get

\begin{equation}
- \frac{\hbar^2}{2m} \frac{d^2 \phi_o}{dx^2} - \frac{D^2}{M a \omega^2_o}
 \phi^3_o = E_o \phi_o
\end{equation}

\indent
This is a non-linear Schr\"{o}dinger equation which can be solved exactly.
There is a localized static solution which reads

\begin{equation}
\phi_o(x) = \frac{\sqrt{g}}{2} sech\left(\frac{g(x-x_o)}{2}\right)
\end{equation}

\noindent
where

\begin{equation}
g = \frac{1}{a} \left(\frac{m}{M}\right) \left(\frac{D}{\hbar\omega_o}\right)^2
\end{equation}

\noindent
and

\begin{equation}
E_o = -\frac{\hbar^2 g^2}{8m}
\end{equation}

\noindent
is the binding energy of the electron in the potential well. $x_o$ is
an arbitrary constant which gives the center of the packet described
by (2.1.11).

\indent
The lattice displacements are given by (2.1.9)

\begin{equation}
\eta_o(x) = -2a\:\left(\frac{\mid E_o \mid}{D}\right)
 sech^2 \left(\frac{g(x-x_o)}{2}\right)
\end{equation}

\noindent
which is symmetric around the electron position.

\indent
The potential well where the electron is trapped is given by:

\begin{equation}
V(x) = -2 \mid E_o \mid sech^2 \left(\frac{g(x-x_o)}{2}\right)
\end{equation}

\indent
We can identify the parameter $g$ as the strength of the interaction
since the potential (2.1.15) becomes very weak for distances greater than
$g^{-1}$. Therefore, $g^{-1}$ defines the polaron length.

\subsection{The Acoustical Case}

\setcounter{equation}{0}

\indent
For electrons interacting with acoustical phonons the Hamiltonian can
be written as \cite{kittel}:

\begin{equation}
H_A = \int{dx} \left\{\frac{\hat{\pi}^2}{2 \nu} + \frac{\nu v^2_s}{2}
\left(\frac{\partial \hat{\eta}}{\partial x}\right)^2
+ \frac{\hbar^2}{2m}\frac{\partial \psi^{\dag}}{\partial x} \frac{\partial
\psi}{\partial x} + D \frac{\partial \hat{\eta}}{\partial x}
\hat{\psi}^{\dag} \psi\right\}
\end{equation}

\noindent
where $v_s$ is the sound velocity in the lattice and all other
definitions are mantained.

\indent
The second term in (2.2.1) comes from the Debye dispersion relation for
acoustical phonons

\begin{equation}
\omega = v_s \mid k \mid
\end{equation}

\noindent
where $k$ is the phonon wave vector.

\indent
For (2.2.1) the ``semiclassical" Lagrangean is

\[
L = \int{dx} \; \{ \frac{\nu}{2} \left(\frac{\partial \eta}{\partial
t}\right)^2 - \frac{\nu v^2_s}{2} \left(\frac{\partial \eta}{\partial
x}\right)^2  + i \hbar \left(\psi^* \frac{\partial \psi}{\partial t} - \psi
\frac{\partial \psi^*}{\partial t}\right)
\]
\begin{equation}
-\frac{\hbar^2}{2m} \frac{\partial \psi^*}{\partial x} \frac{\partial
\psi}{\partial x} - D \frac{\partial \eta}{\partial x} \psi^* \psi \}
\end{equation}

\noindent
which produces the following set of equations of motion

\begin{equation}
i\hbar \frac{\partial \psi}{\partial t} + \frac{\hbar^2}{2m}
\frac{\partial^2 \psi}{\partial x^2} - D \frac{\partial
\eta}{\partial x} \psi = 0
\end{equation}
\begin{equation}
\frac{\partial^2 \eta}{\partial t^2} - v^2_s \frac{\partial^2
\eta}{\partial x^2} - \frac{D}{\nu} \frac{\partial \mid \psi \mid
^2}{\partial x} = 0
\end{equation}

\indent
These equations are interpreted in the same way as for the optical case
as a self consistent interaction, where the potential that the
electron feels is

\[
V(x,t) = D \frac{\partial \eta}{\partial x}
\]

\indent
Unlike (2.1.4) and (2.1.5), this new system of coupled equations
admits a travelling solution.
We can obtain a general solution for (2.2.4) and (2.2.5) defining a new
variable, $\chi = x - x_o - vt$, where $x_o$ and $v$ are arbitrary
constants. We look for solutions of the form

\begin{equation}
\psi(x,t) = \phi_o (\chi) exp\left\{\frac{i}{\hbar} (mvx - E'_o t)\right\}
\end{equation}

\noindent
and

\begin{equation}
\eta(x,t) = \eta_o(\chi)
\end{equation}

\noindent
 From (2.2.4), (2.2.5), (2.2.6) and (2.2.7):

\begin{equation}
-\frac{\hbar^2}{2m} \frac{d^2 \phi_o}{d \chi^2} + D \frac{d \eta}{d \chi}
 = \left( E'_o - \frac{mv^2}{2} \right) \phi_o(\chi)
 \end{equation}
 \begin{equation}
\frac{d \eta_o}{d \chi} = - \frac{D}{\nu(v^2_s - v^2)} \phi^2_o(\chi)
\end{equation}

\indent
Eq.(2.2.8) is the Schr\"{o}dinger equation in the variable $\phi$ and (2.2.9)
has the form of a wave equation of the same variable. Substituting
(2.2.9) in (2.2.8) we get

\begin{equation}
-\frac{\hbar^2}{2m} \frac{d^2 \phi_o}{d \chi^2} - \frac{D^2}{\nu
\left(v^2_s - v^2\right)} \phi^3_o = \left(E'_o - \frac{mv^2}{2}\right) \phi_o
\end{equation}

\indent
Again we have obtained a nonlinear Schr\"{o}dinger equation which is
characteristic of a self consistent interaction. The solution
obtained if $v$ is smaller than the velocity of the sound is

\begin{equation}
\phi_o(\chi) = \frac{\sqrt{g'}}{2} sech\left(\frac{g' \chi}{2}\right)
\end{equation}

\noindent
where

\begin{equation}
g' = \frac{1}{a} \left(\frac{m}{M}\right)
 \left(\frac{D}{\hbar v_s/a}\right)^2 \left(1 -\frac{v^2}{v^2_s}\right)^{-1}
\end{equation}

\noindent
with the electron binding energy given by

\begin{equation}
E'_o = \frac{mv^2}{2} - \frac{\hbar^2 g^{'2}}{8m}
\end{equation}

\indent
The solution is unstable for $v \; > \; v_s$.

\indent
The interpretation is almost obvious: the electron and the lattice
displacement move together with velocity $v$. Observe that the wave
function (2.2.1) and the binding energy (2.2.13) have exactly the same shape
as in the optical case for $v = 0$ (the adiabatic case). We expect
that the potential which the electron feels must be the same. From (2.2.9)

\begin{equation}
\eta_o(\chi) = -\frac{\hbar^2 g'}{2mD}\: tanh\left(\frac{g' \chi}{2}\right)
\end{equation}

\noindent
and the potential is given by

\begin{equation}
V(\chi) = D \frac{d \eta_o}{d \chi} = -\frac{\hbar^2 g^{'2}}{4m}
sech^2\left(\frac{g' \chi}{2}\right)
\end{equation}

\noindent
Observe that for $v = 0$ (2.2.15) has the same shape of (2.1.15). This
exhibits something fundamental in the physics of the two problems.
Observe that the parameters $g$ and $g'$ has the same form for $v = 0$

\[
\frac{1}{a} \left(\frac{m}{M}\right) \left(\frac{D}{E_c}\right)^2
\]

\noindent
where $E_c$ is the characteristic energy of the phonon system; $\hbar
\omega_o$ in the optical case and $\hbar \omega_D$ in the acoustical
case, where

\[
\omega_D = v_s/a
\]

\noindent
is the Debye frequency. So, $g$ and $g'$ play the same role in both
problems and, from now on, we will call them $g$. Actually, we will
work only with the static case, $v = 0$, observing that this means
that we are not taking in account the kinetic energy of the lattice
in (2.2.3) (Born-Oppenheimer approximation). This explains why we are using
the term ``adiabatic".

\indent
Here, it should be noticed that one must reconcile $g >> 1$ (strong coupling)
with the continuum model we have been using. It has been shown in \cite{schu}
that there is a vast range of polaron sizes where both conditions are met.

\subsection{The Adiabatic Expansion}

\setcounter{equation}{0}

\indent
As we have seen, for the acoustical as well as for the optical case,
the potential well which traps the electron is the same. As a
consequence the wave functions and binding energies are also the same.
As we have treated the problem in a variational way we expect that
the wave function, (2.1.11) or (2.2.11), is the ground state wave function
for the electron in the adiabatic limit and zero temperature.
Nevertheless, we expect that for finite temperatures the electron can be
found
in some excited state of this well due to its interaction with
thermal phonons. As the interaction is self consistent the potential
well must change its shape changing the potential energy of the
electron. If the temperature is not too high so it does not remove the
electron from the well, we might imagine that due to virtual transitions,
the electron absorbs energy from the lattice and immediately emits this
energy, remaining in the ground state.

\indent
In order to find these excited states we have to solve the
Schr\"{o}dinger equation for the electron in the potential well, (2.1.15)
or (2.2.15) (put $x_o = 0$)

\begin{equation}
-\frac{\hbar^2}{2m} \frac{d^2}{dx^2} \phi_n(x) -
\frac{\hbar^2g^2}{4m} sech^2 \left(\frac{gx}{2}\right)
 \phi_n(x) = E_n \phi_n(x)
\end{equation}

\indent
This equation can be solved exactly \cite{morse} and gives one bound state
as expected (see (2.1.11) and (2.2.11))

\[
\phi_o(x) = \frac{\sqrt{g}}{2} sech\left(\frac{gx}{2}\right)
\]

\noindent
with energy

\[
E_o = -\frac{\hbar^2 g^2}{8m}
\]

\noindent
and a set of doubly degenerate free states

\begin{equation}
\phi^F_n(x) = \frac{1}{\sqrt{L}}\: e^{ik_nx}\: \left\{\frac{k_n + ig\,
tanh(gx/2)/2}{k_n + ig/2} \right\}
\end{equation}

\noindent
with energy

\begin{equation}
E^F_n = \frac{\hbar^2 k^2_n}{2m} \; \; ,
\end{equation}

\noindent
where $k_n$ is solution of

\begin{equation}
k_n = \frac{2n\pi}{L} - \frac{\delta(k_n)}{L}\;\; n = 0, \pm 1, \pm 2, ...
\end{equation}

\noindent
and $\delta(k)$ is the phase-shift due to the scattering of the free
states given by

\begin{equation}
\delta(k) = arctan \left(\frac{kg}{k^2 - g^2/4}\right)
\end{equation}

\noindent
Here we have imposed periodic boundary condition in $x = \pm L/2$
with $L \rightarrow \infty$.

\indent
The ``adiabatic expansion" is made by expanding the electron field
operators in (2.1.1) or (2.2.1) in this new basis, yielding

\begin{equation}
\hat{\psi}(x,t) = \phi_o(x) \hat{a}_0(t) + \sum^\infty _{n=1}
\phi^F_n(x) \hat{a}_n(t)
\end{equation}
\begin{equation}
\hat{\psi^{\dag}}(x,t) = \phi_o(x) \hat{a}^{\dag}_0(t) + \sum^\infty_{n=1}
\phi^{F*}_n(x) \hat{a}^{\dag} _n(t)
\end{equation}

\noindent
where $a_n$ and $a^{\dag}_n$ are, respectively, the destruction and creation
operators for each state of (2.3.1). They obey the following
anticommutation rules

\[
\left\{\hat{a}_n(t), \hat{a}^{\dag}_m(t)\right\} = \delta_{nm}
\]
\[
\left\{\hat{a}_n(t), \hat{a}_m(t)\right\} =
 \left\{\hat{a}^{\dag}_n(t), \hat{a}^{\dag}_m(t)\right\} = 0
\]

\indent
The field displacement as well as its conjugate momentum density can also
be expanded in this basis as

\begin{equation}
\hat{\eta}(x,t) = \eta_o(x) + \sum^{+ \infty}_{k=- \infty} \hat{q}_k(t)
 \frac{e^{ikx}}{\sqrt{N}}
 \end{equation}
 \begin{equation}
\hat{\pi}(x,t) = \sum^{+ \infty}_{k=- \infty} \hat{p}_k(t)
 \frac{e^{ikx}}{\sqrt{N}}
\end{equation}

\noindent
where $N = L/a$ is the number of ion sites and $k$ is given by the
periodic boundary conditions

\[
k = \frac{2n\pi}{L}\;\; n = 0, \pm 1, \pm 2, ...
\]

\noindent
As $\hat{\eta}$ and $\hat{\pi}$ are real we must have

\[
\hat{q}^{\dag}_k = \hat{q}_{-k}
\]
\[
\hat{p}^{\dag}_k = \hat{p}_{-k}
\]

\noindent
with the usual commutation rule

\[
[\hat{q}_k(t), \hat{p}_{k'}(t)] = i\hbar \delta_{k,k'}
\]

\indent
If we substitute (2.3.6), (2.3.7), (2.3.8) and (2.3.9) in (2.1.1) and (2.2.1),
 and use the orthogonality of the adiabatic states

\begin{equation}
\int{dx} \phi_o(x) \phi^F_n(x) = 0
\end{equation}
\begin{equation}
\int{dx} \phi^F_n(x) \phi^F_{n'}(x) = \delta_{nn'}
\end{equation}

\noindent
and the equations (2.1.9), (2.2.9) and (2.3.1) we obtain

\indent
i) for the optical case:

\[
H_o = \epsilon_o + E_o a^{\dag}a + \sum^{\infty}_{n=0} E^F_n a^{\dag}_n a_n +
\]
\[
\sum^{+\infty}_{k=-\infty} \{ \frac{\hat{p}_k \hat{p}_{-k}}{2M} +
\frac{M \omega^2_o}{2} \hat{q}_k \hat{q}_{-k}
+ \frac{D}{a} q_k [ f^o(k)(a^{\dag}a-1) +
\]
\begin{equation}
\sum^\infty_{n=0} ( f_n(k) a^{\dag} a_n + f^*_n(-k) a^{\dag}_n a +
\sum^\infty_{m=0} f_{nm}(k)a^{\dag}_n a_m ) ] \}
\end{equation}

\noindent
where

\begin{equation}
\epsilon_o = \frac{\nu \omega^2_o}{2} \int^{+\infty}_{-\infty} dx
\eta^2_o(x) = \frac{\hbar^2g^2}{12m} = \frac{2}{3} \mid E_o \mid
\end{equation}
\begin{equation}
f^o(k) = \frac{1}{\sqrt{N}} \int{dx} \phi^2_o(x) e^{ikx}
\end{equation}
\begin{equation}
f_n(k) = \frac{1}{\sqrt{N}} \int{dx} \phi_o(x) \phi^F_n(x) e^{ikx}
\end{equation}
\begin{equation}
f_{nm}(k) = \frac{1}{\sqrt{N}} \int{dx} \phi^{F*}_n(x) \phi^F_m(x) e^{ikx}
\end{equation}

\noindent
ii) for the acoustical case:

\newpage

\[
H_A = \epsilon_o + E_oa^{\dag}a + \sum^\infty_{n=o} E^F_n a^{\dag}_n a_n +
\]
\[
\sum^{+\infty}_{k=-\infty} \{ \frac{\hat{p}_k \hat{p}_{-k}}{2M} +
\frac{Mv^2_s k^2}{2} \hat{q}_k \hat{q}_{-k}
+ ikDq_k [f_o(k) (a^{\dag}a-1)
\]
\begin{equation}
+ \sum^\infty_{n=0} (f_n(k)a^{\dag}a_n + f^*_n(-k)a^{\dag}_n a
+ \sum^\infty_{m=0} f_{nm}(k) a^{\dag}_n a_m )] \}
\end{equation}

\noindent
Notice that (2.3.12) and (2.3.17) are exact results, no approximations
have been made so far.

\indent
Since we are interested in the strong coupling limit, the phonons can not
excite the electron out of the well. Quantitatively this means that

\[
\frac{\mid E_o \mid}{E_c} \gg 1
\]

\noindent
where $E_c$ is the phonon characteristic energy and $\mid
E_o \mid$ the modulus of the binding energy of the electron. The
inequality above allows us to define an expansion parameter for
the strong coupling limit, $\alpha$, as

\[
\alpha = \frac{E_c}{\mid E_o \mid} \ll 1
\]

\indent
If now we scale the Hamiltonians (2.3.12) and (2.3.17) by the characteristic
phonon energy we can easily see that the pure electronic part is of the
order of $\alpha^{-1}$ while the interaction term is of the order of
$\alpha^{-1/2}$ and consequently much smaller than the former \cite{shaw},
\cite{schu}. Using
this result we will eliminate the electronic part of the Hamiltonian by a
perturbative treatment and obtain a renormalized phonon Hamiltonian.

\indent
Consider the pure electronic part as the non-perturbed Hamiltonian.
Its eigenstates can be written in the Fock space as:

\begin{equation}
\mid  n_o, n_1, ... n_\infty >
\end{equation}

\noindent
where $n_o$ is the occupation number for the ground state. We
simplify the calculation assuming that there is only one electron in
the problem (see appendix A for the case of many electrons), that is:

\begin{equation}
\sum^\infty_{j=0} n_j = 1
\end{equation}

\indent
The ground state is given by:

\begin{equation}
\mid \psi_o > = \mid n_o = 1 ; n_j = 0 > \, \,  j = 1,2,3,...
\end{equation}

\noindent
with energy $E_o$.

\indent
This state is exactly the adiabatic state given by (2.1.11) or (2.2.11).

\indent
We can now apply the Rayleigh-Schr\"{o}dinger perturbation theory to this
problem. The first order term is null because we have just one electron. The
second order correction is easily calculated and it gives,

\indent
i) for the optical case:

\begin{equation}
E^{(2)}_0 = -\frac{D^2}{a^2} \sum_{kk'} V_o(k,k') \hat{q}_k \hat{q}_{-k'}
\end{equation}

\noindent
where:

\begin{equation}
V_o(k,k') = \sum^\infty_{n=0} \frac{f_n(k) f^*_n(k')}{E^F_n - E_o}
\end{equation}

\noindent
ii) for the acoustical case:

\begin{equation}
E^{(2)}_A = -D^2 \sum_{kk'} kk' V_o(k,k') \hat{q}_k \hat{q}_{-k'}
\end{equation}

\noindent
Therefore, we write the Hamiltonians for the renormalized phonons as

\begin{equation}
H_{O,A} = \epsilon + \sum^{+\infty}_{k=-\infty} \{ \frac{\hat{p}_k
\hat{p}_{-k}}{2M} + \frac{M}{2}
\sum^{+\infty}_{k'=-\infty} \left[\Omega^{O,A}_{kk'}\right]^2 \hat{q}_k
\hat{q}_{-k'} \}
\end{equation}

\noindent
where

\begin{equation}
\epsilon = \epsilon_o + E_o = - \frac{\mid E_o \mid}{3}
\end{equation}

\noindent
and

\begin{equation}
\left( \Omega^{O}_{kk'} \right)^2 = \omega^2_o \delta_{kk'} -\frac{2D^2}{Ma^2}
V_o(k,k')
\end{equation}
\begin{equation}
\left( \Omega^{A}_{kk'}\right)^2 = v^2_sk^2 \delta_{kk'} - \frac{2D^2}{M} kk'
 V_o(k,k')
\end{equation}

\indent
The Hamiltonian (2.3.24) was already obtained \cite{shaw} some years ago for
the acoustical case and it can be rewritten in a different form in terms of
some new operators $\delta \hat{\eta}(x,t)$ and $\hat{\pi}(x,t)$ defined by
(c.f. (2.3.8))

\begin{equation}
\hat{q}_k(t) = \frac{1}{\sqrt{N}} \int{\frac{dx}{a}} e^{-ikx}
\delta \hat{\eta}(x,t)
\end{equation}
\begin{equation}
\hat{p}_k(t) = \frac{1}{\sqrt{N}} \int{dx} e^{-ikx} \hat{\pi}(x,t)
\end{equation}

\noindent
One can easily check that

\[
[ \delta \hat{\eta}(x,t) , \hat{\pi}(x',t)] = i\hbar \delta(x-x')
\]

\noindent
Now, using (2.3.28) and (2.3.29) in (2.3.24) we get

\begin{equation}
H_O = \epsilon + \int{dx} \left\{ \frac{\hat{\pi}^2}{2\nu} + \frac{\nu
\omega^2_o}{2} \delta \hat{\eta}^2
- \frac{\nu \omega^2_o}{2} \delta \hat{\eta} \int{dx'} F(x,x')
\delta \hat{\eta}(x')
 \right\}
\end{equation}

\noindent
and

\begin{equation}
H_A = \epsilon + \int{dx} \left\{ \frac{\hat{\pi}^2}{2\nu} + \frac{\nu
v^2_s}{2} \left(\frac{\partial \delta \hat{\eta}}{\partial x}\right)^2
- \frac{\nu v^2_s}{2} \frac{\partial \delta \hat{\eta}}{\partial x} \int{dx'}
F(x,x') \frac{\partial \delta \hat{\eta}}{\partial x'} \right\}
\end{equation}

\noindent
where

\begin{equation}
F(x,x') = 4g\; \phi_o(x) \phi_o(x') \sum^\infty_{n=0}
\frac{\phi^{F *}_n(x) \phi^F_n(x')}{\left(k^2_n + g^2/4\right)}
\end{equation}

\noindent
These results agree perfectly with those obtained by other methods
\cite{schu}, \cite{holstein}.

\renewcommand{\theequation}{\thesection.\arabic{equation}}
\setcounter{equation}{0}

\section{The Effective Hamiltonian}

\indent
As we are interested in the excitation spectrum of the Hamiltonians (2.3.30)
and (2.3.31), we have to diagonalize them.  If we choose

\begin{equation}
\delta \hat{\eta}(x,t) = \sum^\infty_{n=0} \hat{q}_n(t) u_n(x) ,
\end{equation}

\noindent
where $u_n$ are the normalized eigenfunctions of (2.3.30) or (2.3.31)

\begin{equation}
\int{\frac{dx}{a}} u^*_n(x) u_m(x) = \delta_{nm} ,
\end{equation}

\noindent
they must satisfy the following integro-differential equations

\begin{equation}
i) \; \; \Omega^2_n u_n(x) = \omega^2_o u_n(x)-\omega^2_o \int{dx'}
F(x,x')u_n(x
   ')
\end{equation}

\noindent
for the optical case, while

\begin{equation}
ii) \; \; \Omega^2_n u_n(x) = -v^2_s \frac{d^2 u_n}{dx^2}
+ v^2_s \int{dx'} \frac{\partial}{\partial x} F(x,x')
\frac{du_n(x')}{dx'}
\end{equation}

\noindent
for the acoustical case.

\indent
Observe that to solve (3.3) and (3.4) is equivalent in the momentum
space to diagonalize (2.3.26) and (2.3.27).

\indent
These equations often appear  in works about polaron dynamics
\cite{schu}, \cite{hols}, \cite{turke} and we do not intend to
solve them in this article. For (3.3) there is a closed solution
\cite{turke} which we will use later, while for (3.4) we will make some
approximations which are suitable for our purposes.

\indent
An interesting and important solution of those equations can be
found directly. These are the zero mode solutions, that is,
solutions with $\Omega_o = 0$ (see appendix B). For the optical case
we have

\begin{equation}
u^{OP}_o(x) = \frac{\sqrt{15ag}}{2} tanh \left(\frac{gx}{2}\right)
 sech^2 \left(\frac{gx}{2}\right)
\end{equation}

\noindent
and for the acoustical case

\begin{equation}
u^{AC}_o(x) = \sqrt{\frac{3ag}{8}} sech^2 \left(\frac{gx}{2}\right)
\end{equation}

\indent
Examining (3.5) and (3.6) and comparing with (2.1.14) and (2.2.14) we see that,
in both cases,

\begin{equation}
u_o(x) = C \; \frac{d}{dx} \eta_o(x)
\end{equation}

\noindent
where C is a constant which appears due to the normalization of $u_o$.
The above relation clearly expresses the translational invariance of
Hamiltonians (2.1.1) and (2.2.1). Notice that although we have put $x_o =
0$ in (2.3.1), in order to obtain the adiabatic basis, the
center of the soliton solutions (2.1.11), (2.1.14), (2.2.11) and (2.2.14), is
arbitrary and therefore we must have the $u_n$ also expanded about this
point, that is:

\begin{equation}
u_n \equiv u_n(x-x_o)
\end{equation}

\indent
Suppose that we move the center of the functions (2.1.14) or (2.2.14) by an
infinitesimal quantity $\delta x_o$.Then,

\begin{equation}
\eta_o(x_o+\delta x_o) \simeq \eta_o(x_o) + \frac{u_o(x_o) \delta x_o}{C}
\end{equation}

\noindent
where we have used (3.7). By (3.9) we conclude that the zero mode frequency
corresponds to the translation of the soliton, in other words,
to the motion of the polaron.

\indent
Once we have the eigenfunctions of (3.3) and (3.4), we would expect to
write the Hamiltonian in the form:

\begin{equation}
H = \epsilon + \frac{\hat{P}^2_o}{2M} + \sum^\infty_{n=1}
\left(\frac{\hat{P}^2_n}{2M} + \frac{M\Omega^2_n}{2} \hat{q}^2_n\right)
\end{equation}

\noindent
where we have used that $\Omega_o = 0$.

\indent
At first sight (3.10) shows a free particle with momentum $P_o$ and a
set of decoupled harmonic oscillators. Nevertheless, it is not
possible to take it seriously because, initially, we have implicitly assumed
that the lattice displacement can not be indefinitely large. From (2.3.21)
and (2.3.23) we see that the energy correction depends on the lattice
displaceme
   nt
which must be finite in order to the pertubation theory to be valid.

\indent
Let us observe that, due to (3.9), the polaron displacement is
proportional to the displacement of its center, that is:

\begin{equation}
q_o = \delta x_o/C
\end{equation}

\noindent
 From (3.1), we have

\[
\delta \eta(x,t) = q_o u_o + \sum^\infty_{n=1} q_n u_n
\]

\noindent
or using (3.7) and (3.11)

\begin{equation}
\delta \eta(x,t) = \delta x_o \frac{\partial \eta_o}{\partial x_o} +
\sum^\infty_{n=1} q_n u_n
\end{equation}

\noindent
Therefore we can assumed that $x_o$ is a true dynamic variable, that
is, $x_o = x_o(t)$. So, based on (3.12), we will rewrite the expansion
(3.1) as:

\begin{equation}
\delta \hat{\eta}(x,t) = \eta_o(x-\hat{x}_o(t)) + \sum^\infty_{n=1}
\hat{q}_n(t) u_n(x-\hat{x}_o(t))
\end{equation}

\indent
This procedure is known as ``collective coordinate formalism"
\cite{raja}. Observe that (3.13) changes the kinetic part of (2.3.30) or
(2.3.31) because $\hat{x}_o$ is also a function of time. It is shown
\cite{schu} that the new Hamiltonian in the presence of the
polaron position operator, $\hat{x}_o$, is given by

\newpage

\[
H = \epsilon + \sum^\infty_{n=1} \left( \frac{\hat{p}^2_n}{2M} +
\frac{M\Omega^2_n}{2} \hat{q}^2_n \right) +
\]
\[
\frac{1}{8M_o} \left\{\left( \hat{P} - \sum^\infty_{n,m=1} G_{nm} \hat{q}_n
 \hat{p}_n \right),
\left(1 + \sum^\infty_{n=1} S_n \hat{q}_n \right)^{-1}\right\}
\]
\begin{equation}
- \frac{\hbar^2 M}{8M^2_o} \sum^\infty_{n=1} S^2_n \left( 1 +
\sum^\infty_{m=1} S_m \hat{q}_m \right)^{-1}
\end{equation}

\noindent
here $\{,\}$ denotes anticommutation and
$\hat{P}$ is the momentum operator associate to $\hat{x}_o$

\[
[\hat{x}_o(t), \hat{P}(t)] = i\hbar
\]

\noindent
and

\[
[\hat{q}_n(t), \hat{p}_m(t)] = i\hbar \delta_{nm}
\]

\noindent
with all the other commutators being zero.

\indent
In (3.14) we have

\begin{equation}
M_o = \nu \int^{+\infty}_{-\infty}{dx} \left[\frac{d\eta_o}{dx}\right]^2
\end{equation}

\noindent
as being the classical soliton mass which becomes

\[
M_o = \frac{m}{8} \left(\frac{E_o}{\hbar \omega_o}\right)^2
\]

\noindent
for the optical case, and

\[
M_o = \frac{32}{3} m \left(\frac{E_o}{\hbar v_s g}\right)^2
\]

\noindent
for the acoustical case.

\indent
The new quantities

\begin{equation}
S_n = \frac{M}{M_o} \int{\frac{dx}{a} \frac{d\eta_o}{dx} \frac{du_n}{dx}}
\end{equation}

\noindent
and

\begin{equation}
G_{nm} = \int{\frac{dx}{a} u_m(x) \frac{du_n(x)}{dx}}
\end{equation}

\noindent
couples the polaron to the renormalized phonons.

\indent
In the strong coupling limit (3.14) can be simplified \cite{hols}. Due to
(3.5), (3.6), (2.1.14), (2.2.14) and (3.7), we can rewrite $S_n$ as

\[
S_n \sim \frac{M}{M_o} \frac{1}{\sqrt{ag}} \left( \frac{E_o}{D} \right)
\int{\frac{dx}{a}} u_o(x) \frac{du_n}{dx}(x)
\]

\noindent
As the integral only gives a numerical factor, this yields

\[
S_n \sim \frac{M}{M_o} \frac{1}{\sqrt{ag}} \left(\frac{E_o}{D}\right)
 \frac{1}{a}
\]

\noindent
Now, from (3.15),

\[
\frac{M}{M_o} \sim \frac{M}{m} \left(\frac{E_c}{E_o}\right)^2
\]

\noindent
and

\[
q_n \sim \frac{\hbar}{(M E_c)^1/2}
\]

\noindent
so

\[
\sum_{n} S_n q_n \sim \left[\frac{E_c}{\mid E_o \mid} \right]^{1/2}
\ll 1
\]

\noindent
and therefore, this sum is very small in the strong coupling limit. Within this
approximation we get

\begin{equation}
H \simeq \epsilon + \sum^\infty_{n=1} \left( \frac{\hat{p}^2_n}{2M} +
\frac{M \Omega^2_n}{2} \hat{q}^2_n \right) +
\frac{1}{2M_o} \left(\hat{P} - \sum^\infty_{n,m=1} G_{nm} \hat{q}_m
\hat{p}_n \right)^2
\end{equation}

\indent
The second term in (3.18) is the energy of non-interacting phonons and
the third term can be interpreted as the kinetic energy of the
polaron. Observe that

\begin{equation}
\dot{\hat{x}_o} = \frac{1}{i \hbar} [\hat{x}_o, \hat{H} ] =
\frac{1}{M_o} \left(\hat{P} - \sum^\infty_{n,m=1} G_{nm} \hat{q}_m
\hat{p}_n \right)
\end{equation}

\indent
and so, $\hat{P}$ can not be the polaron momentum because since

\[
\dot{\hat{P}} = \frac{1}{i\hbar} [\hat{P}, \hat{H}] = 0
\]

\noindent
it is a constant of motion. From (3.19) we interpret $M_o\dot{\hat{x}_o}$ as
the polaron momentum and $\sum_{nm} G_{nm}\hat{q}_m \hat{p}_n$ the
momentum of the phonon field. Observe that Hamiltonian (3.18) is very close to
the electromagnetic Hamiltonian, where the coupling between the particle
and the field are obtained via the potential vector (see Section 4).

\indent
If we define the destruction and creation operators

\begin{equation}
\hat{b}_n = \left( \frac{M \Omega_n}{2 \hbar} \right)^{1/2} \left(
 \hat{q}_n + i \frac{\hat{p}_n}{M \Omega_n} \right)
\end{equation}
\begin{equation}
\hat{b}^{\dag}_n = \left( \frac{M \Omega_n}{2 \hbar} \right)^{1/2}
 \left( \hat{q}_n - i \frac{\hat{p}_n}{M \Omega_n} \right)
\end{equation}

\noindent
which obviously obey

\[
[\hat{b}_n, \hat{b}^{\dag}_m] = \delta_{nm}
\]

\noindent
one can rewrite H as

\begin{equation}
H = \epsilon + \sum^\infty_{n=1} \hbar \Omega_n(b^{\dag}_nb_n + 1/2) +
+ \frac{1}{2M_o} (\hat{P} - \hat{P}_{ph})^2
\end{equation}

\noindent
where

\[
\hat{P}_{ph} = \sum^\infty _{n,m=1} \frac{\hbar}{2i}
\left[\left(\frac{\Omega_n}{\Omega_m}\right)^{1/2} + \left(\frac{\Omega_m}
{\Omega_n}\right)^{1/2}\right] G_{nm}b^{\dag}_mb_n
\]
\begin{equation}
+ \sum^\infty_{n,m=1} \frac{\hbar}{4i}
\left[\left(\frac{\Omega_n}{\Omega_m}\right)^{1/2} -
\left(\frac{\Omega_m}{\Omega_n}\right)^{1/2}\right]
 G_{nm}\left(b_mb_n - b^{\dag}_mb^{\dag}_n\right)
\end{equation}

\noindent
Here we have used the fact that, from (3.15), $G_{nm}$ is anti-symmetric in
the interchange of m and n,

\begin{equation}
G_{nm} = -G_{mn}
\end{equation}

\indent
Observe that the momentum of the phonon field consists on two parts;
a diagonal part (the first term on the right hand side of (3.23)) which
commutes with the phonon-number operator:

\[
\hat{N} = \sum^\infty_{n=1} \hat{b}^{\dag}_n \hat{b}_n
\]

\noindent
and, therefore, conserves the number of phonons in the system. This
term is responsible for scattering. The other term does not commute
with the number operator and is related with absorption or emission
of phonons by the polaron (Cerenkov process). We will restrict our
problem to typical polaron kinetic energies much smaller than the
phonon energies, in other words, small velocities. In this limit
the occurrence of emission or absorption of phonons is not possible
due to the simultaneous conservation of momentum and energy. In terms
of our parameters this means that:

\begin{equation}
\mid \dot{x}_o \mid \ll \sqrt{\frac{E_c}{M_o}}
\end{equation}

\noindent
Only scattering, and therefore virtual transitions, will be relevant
for our problem.

\indent
With this approximation the polaron dynamics will be described by the
following effective Hamiltonian:

\begin{equation}
H = \frac{1}{2M_o} \left(\hat{P} - \sum^\infty_{n,m=1}
\hbar g_{nm} \hat{b}^{\dag}_m \hat{b}_n\right)^2
+ \sum^\infty_{n=1} \hbar \Omega_n \hat{b}^{\dag}_n \hat{b}_n
\end{equation}

\noindent
where

\begin{equation}
g_{nm} = \frac{1}{2i} \frac{(\Omega_n + \Omega_m)}{\sqrt{\Omega_n
\Omega_m}} G_{nm}
\end{equation}

\indent
As we will show in the next section the Hamiltonian (3.26) describes
the dynamics of a brownian particle, that is, a heavy particle in a
bath of light particles which collide with it.

\setcounter{equation}{0}

\section{Functional Integral Method}

\indent
The starting point for the calculations of the transport properties
of the polaron is the well-known Feynman-Vernon formalism \cite{feynman}
that the authors have recently applied \cite{casca} to the Hamiltonian
(3.26).

\indent
We are interested only in the quantum statistical properties of the polaron
and the phonons act only as a source of relaxation and diffusion
processes. Consider the density operator for the system polaron plus
phonons, $\hat{\rho}(t)$. This operator evolves in time according to

\begin{equation}
\hat{\rho}(t) = e^{-i\hat{H}t/\hbar} \hat{\rho}(0) e^{i\hat{H}t/\hbar}
\end{equation}

\noindent
where $\hat{H}$ is given by (3.26) and $\hat{\rho}(0)$ is the density
operator at $t = 0$ which we will assume to be decoupled as a product
of the polaron density operator, $\hat{\rho}_S(0)$, and the phonon
density operator, $\hat{\rho}_R(0)$:

\begin{equation}
\hat{\rho}(0) = \hat{\rho}_S(0) \hat{\rho}_R(0)
\end{equation}

\noindent
where the symbol $\underline{S}$ refers to the polaron (system of
interest) and $\underline{R}$ to the phonons (the reservoir of excitations).

\indent
The condition (4.2) means that we put the electron
in the lattice which is in thermal equilibrium at temperature T.
So, we consider the phonons as described by their equilibrium
distribution,

\begin{equation}
\hat{\rho}_R(0) = \frac{e^{-\beta \hat{H}_R}}{Z}
\end{equation}

\noindent
where

\begin{equation}
Z = tr_R \left(e^{-\beta \hat{H}_R}\right)
\end{equation}

\noindent
with

\begin{equation}
\beta = \frac{1}{K_BT}
\end{equation}

\noindent
Here $tr_R$ denotes the trace over the phonons variables and $K_B$ is
the Boltzmann constant. $\hat{H}_B$ is the free phonon Hamiltonian
which is given by the last term on the right hand side of (3.26).

\indent
As we said, we are interested only in the quantum dynamics of the system S,
so, we define a reduced density operator

\begin{equation}
\hat{\rho}_s(t) = tr_R (\hat{\rho}(t))
\end{equation}

\noindent
which contains all the information about S when it is in contact with R.

\indent
Projecting now (4.6) in the coordinate representation of the polaron system

\begin{equation}
\hat{x}_o \mid q > = q \mid q >
\end{equation}

\noindent
and in the coherent state representation for bosons (the phonons)

\begin{equation}
\hat{b}_n \mid \alpha_n > = \alpha_n \mid \alpha_n >
\end{equation}

\noindent
we have \cite{castro neto} (see also appendix C)

\begin{equation}
\rho_s (x, y, t) = \int{dx'} \int{dy'} J(x, y, t; x', y',0) \rho_s(x',y',0)
\end{equation}

\indent
Here we have used (4.1),(4.2),(4.6) and the completeness relation for the
representations above, namely,

\begin{equation}
\int{dq} \mid q > < q \mid = 1
\end{equation}
\begin{equation}
\int{\frac{d^2 \alpha}{\pi}} \mid \alpha > < \alpha \mid = 1
\end{equation}

\noindent
where $d^2 \alpha = d(Re \alpha)d(Im \alpha)$ as usual.

\indent
In (4.9), $J$ is the superpropagator of the polaron, which can be written
as

\begin{equation}
J = \int^{x}_{x'}{Dx} \int^{y}_{y'}{Dy} \, \,
e^{\frac{i}{\hbar}(S_o[x]-S_o[y])} F[x,y]
\end{equation}

\noindent
where

\begin{equation}
S_o[x] = \int^t_0{dt'} \; \left\{\frac{M_o \dot{x}^2(t')}{2}\right\}
\end{equation}

\noindent
is the classical action for the free particle. F is the so-called
influence functional

\[
F[x,y] = \int{\frac{d^2 \vec{\alpha}}{\pi^N}} \int{\frac{d^2
\vec{\beta}}{\pi^N}} \int{\frac{d^2 \vec{\beta}'}{\pi^N}} \;
\rho_R(\vec{\beta}^*, \vec{\beta}') \; e^{-\mid \vec{\alpha} \mid^2 -
\frac{\mid \vec{\beta} \mid^2}{2} - \frac{\mid \vec{\beta}'\mid^2}{2}}
\]
\begin{equation}
\int^{\vec{\alpha}^*}_{\vec{\beta}} {D^2 \vec{\alpha}}
\int^{\vec{\alpha}}_{\vec{\beta}'^*} {D^2 \vec{\gamma}} \; e^{S_I[x,
\vec{\alpha}] + S^*_I[y,\vec{\gamma}]}
\end{equation}

\noindent
where $\vec{\beta}$ denotes the vector $(\beta_1, \beta_2,
\beta_3,..., \beta_N)$ and $S_I$ is a complex action related to the
reservoir plus interaction

\begin{equation}
S_I[x,\vec{\alpha}] = \int^t_0{dt'} \left\{ \frac{1}{2}
\left( \vec{\alpha} \cdot
\frac{d \vec{\alpha}^*}{dt'} - \vec{\alpha}^* \cdot
\frac{d\vec{\alpha}}{dt'}\right)
- \frac{i}{\hbar} (H_R - \dot{x} h_I) \right\}
\end{equation}

\noindent
with

\begin{equation}
H_R = \sum^\infty_{n=1} \hbar \Omega_n \alpha^*_n \alpha_n
\end{equation}
\begin{equation}
h_I = \sum^\infty_{n,m=1} \hbar g_{nm} \alpha^*_m \alpha_n
\end{equation}

\noindent
Here we have obtained a result which is very close to the
electromagnetic coupling where the Hamiltonian depends on the
vector potential, $\vec{A}$, through

\[
\left(\vec{p} - \frac{e \vec{A}}{c}\right)^2
\]

\noindent
but the Lagrangean depends on

\[
\vec{v} \cdot \vec{A}
\]

\indent
In our case the Lagrangean formulation simplifies the problem
transforming a nonlinear problem into a linear one. The action (4.15) is
quadratic in $\vec{\alpha}$, so it can be solved exactly. Observe
that the Euler-Lagrange equations for (4.15) are

\begin{equation}
\dot{\alpha}_n + i\Omega_n \alpha_n - i \dot{x} \sum^\infty_{m=1}
g_{mn}\alpha_m = 0
\end{equation}
\begin{equation}
\dot{\alpha}^*_n - i \Omega_n \alpha^*_n + i \dot{x}
\sum^\infty_{m=1} g_{nm} \alpha^*_m = 0
\end{equation}

\noindent
which must be solved subject to the boundary conditions

\begin{equation}
\alpha_n(0) = \beta_n
\end{equation}
\begin{equation}
\alpha^*_n(t) = \alpha^*_n
\end{equation}

\indent
Due to (3.24) we have $g_{nn} = 0$, so, the modes are not coupled among
themselves. This makes (4.18) and (4.19) easy to solve. That set of
equations represents a set of harmonic oscillators forced by the
presence of the polaron. The result can be written as:

\begin{equation}
\alpha_n(\tau) = e^{-i \Omega_n \tau} \left(\beta_n + \sum^\infty_{m=1}
W_{nm}(\tau) \beta_m \right)
\end{equation}
\begin{equation}
\alpha^*_n(\tau) = e^{i \Omega_n\tau} \left(\alpha^*_n e^{-i \Omega_nt}
 +\sum^\infty_{m=1} \tilde{W}_{nm}(\tau) e^{-i \Omega_m t} \alpha^*_m \right)
\end{equation}

\noindent
where $W_{nm}$ and $\tilde{W}_{nm}$ are functionals of $x(t)$
which obey the following equations

\begin{equation}
W_{nm} = W^o_{nm} + \sum^\infty_{n'=1} W^o_{nn'} W_{n'm}
\end{equation}
\begin{equation}
\tilde{W}_{nm} = \tilde{W}^o_{nm} + \sum^\infty_{n'=1}
\tilde{W}^o_{nn'} \tilde{W}_{n'm}
\end{equation}

\noindent
where

\begin{equation}
W^o_{nm}([x], \tau) =
  i \int^\tau_0{dt'} g_{nm} \dot{x}(t') e^{i(\Omega_n - \Omega_m)t'}
  \end{equation}
  \begin{equation}
\tilde{W}^o_{nm}([x], \tau) =
 i \int^t_\tau{dt'} g_{mn} \dot{x}(t') e^{i(\Omega_m - \Omega_n)t'}
\end{equation}

\noindent
(observe that: $W_{nm}(t) = \tilde{W}_{mn}(0)$).

\indent
Now we expand the action (4.15) around the classical solution (4.22)
and (4.23) and obtain, after some integrations in (4.14)

\begin{equation}
F[x,y] = \prod^\infty_{n=1} (1-\Gamma_{nn}[x,y] \; \overline{n}_n)^{-1}
\end{equation}

\noindent
where

\begin{equation}
\Gamma_{nm} = W^*_{nm}[y] + W_{mn}[x] + \sum^\infty_{\ell=1}
W^*_{\ell m}[y] W_{\ell n}[x]
\end{equation}

\noindent
with

\begin{equation}
\overline{n}_n = \left(e^{\beta \hbar \Omega_n} - 1\right)^{-1}
\end{equation}

\indent
Notice that (4.28) and (4.29) are exact, no approximations have been
made so far.

\indent
We see from (4.24) and (4.25) that $W_{nm}$ can be expressed as a power
series of the Fourier transform of the polaron velocity,
$\dot{x}$, so, due to the small polaron velocity condition
(3.25), we expect that only few terms in (4.24) will be sufficient for a
good description of the polaron dynamics.

\indent
Another way to see this, is to notice that (4.24) and (4.25) are the
scattering amplitudes from the mode $k$ to the mode $j$. The terms that appear
in the sum represent the virtual transitions between these two modes.
With these two arguments in mind we will make use of the
Born-approximation. In matrix notation,

\begin{equation}
W = (1-W^o)^{-1} W^o \simeq W^o + W^oW^o
\end{equation}

\indent
Therefore, in the approximation of small polaron velocity
the terms in (4.29) are small and we can rewrite as a good approximation

\begin{equation}
F[x,y] \simeq exp\left\{\sum^\infty_{n=1} \Gamma_{nn}[x,y] \;
\overline{n}_n\right\}
\end{equation}

\indent
Observe that if the interaction is turned off $(\Gamma \rightarrow
0)$ or the temperature is zero $(T = 0)$ the functional (4.29) is one,
and, as we would expect the polaron moves as a free particle.

\indent
Substituting the Born approximation (4.31) in (4.32) and the latter in (4.12)
we find

\begin{equation}
J = \int^{x}_{x'}{Dx} \int^{y}_{y'}{Dy}\, exp\left\{\frac{i}{\hbar}
\tilde{S}[x,y] + \frac{1}{\hbar} \tilde{\phi}[x,y]\right\}
\end{equation}

\noindent
where

\begin{equation}
\tilde{S} = \int^t_0{dt'} \left\{ \frac{M_o}{2}(\dot{x}^2(t') - \dot{y}^2(t'))
+(\dot{x}(t') - \dot{y}(t')) \int^t_0{dt"} \; \Gamma_I(t'-t") (\dot{x}(t") +
\dot{y}(t"))\right\}
\end{equation}

\noindent
and

\begin{equation}
\tilde{\phi} = \int^t_0{dt'} \; \int^t_0{dt"} \left\{\Gamma_R(t'-t")
 (\dot{x}(t') - \dot{y}(t'))(\dot{x}(t") - \dot{y}(t"))\right\}
\end{equation}

\noindent
with

\begin{equation}
\Gamma_R(t) = \hbar \theta(t) \sum^\infty_{n,m=1} g^2_{nm}
\overline{n}_n cos(\Omega_n - \Omega_m)t
\end{equation}
\begin{equation}
\Gamma_I(t) = \hbar \theta(t) \sum^\infty_{n,m=1} g^2_{nm}
\overline{n}_n sin(\Omega_n - \Omega_m)t
\end{equation}

\indent
Now, if we define the new variables R and r as

\begin{equation}
R = \frac{x + y}{2}
\end{equation}
\begin{equation}
r = x - y
\end{equation}

\noindent
the equations of motion for the action in (4.34) read

\begin{equation}
\ddot{R}(\tau) + 2 \int^t_0{dt'} \gamma(\tau-t') \dot{R}(t') = 0
\end{equation}
\begin{equation}
\ddot{r}(\tau) - 2 \int^t_0{dt'} \gamma(t'-\tau) \dot{r}(t') = 0
\end{equation}

\noindent
where

\begin{equation}
\gamma(t) = \frac{1}{M_o} \frac{d\Gamma_I}{dt}
\end{equation}

\noindent
or, using (4.37)

\begin{equation}
\gamma(t) = \frac{\hbar \theta(t)}{M_o} \sum^\infty_{n,m=1}
g^2_{nm} \overline{n}_n(\Omega_n - \Omega_m) cos(\Omega_n - \Omega_m)t
\end{equation}

\noindent
is the damping function.

\indent
In terms of these newly defined variables, we can easily see that
(4.40) and (4.41) have the same form of the equations previously obtained
in the case of quantum brownian motion \cite{caldeira}, except for the
fact that they now present memory effects. It should be emphasized that
although (4.40) and (4.41) have only indirect physical meaning, through
the study of the motion of the center of a wavepacket and the spreading
of its width, $\gamma(t)$ really plays the role of the damping parameter
in the equation of motion of the former (see ref. \cite{caldeira} for
details).

\indent
Furthermore, we shall prove that (4.43) can be written in the form

\begin{equation}
\gamma(t) = \overline{\gamma}(T) \delta(t)
\end{equation}

\noindent
where $\overline{\gamma}(T)$ is a damping parameter which is
temperature dependent and $\delta(t)$ is the Dirac delta function.
The form (4.44) is known as the Markovian approximation, because in this case
the memory is purely local and does not depend on the previous motion
of the particle.

\indent
If we use (4.40) and (4.41) with (4.44) and expand the phase of (4.33)
around this classical solution we get the well-known result for the
quantum brownian motion \cite{caldeira} where the damping parameter
$\gamma$ (temperature independent) is replaced by
$\overline{\gamma}(T)$ and the diffusive part is replaced by (4.35).
As a consequence, the diffusion parameter in momentum space will be
given by

\begin{equation}
D(t) = \hbar \frac{d^2 \Gamma_R}{dt^2} = -\hbar^2\theta(t)
 \sum^\infty_{n,m=1} g^2_{nm} \overline{n}_n(\Omega_n - \Omega_m)^2
cos(\Omega_n - \Omega_m)t
\end{equation}

\indent
We will also prove that $D(t)$ has the Markovian form:

\begin{equation}
D(t) = \overline{D}(T) \delta(t)
\end{equation}

\noindent
where $\overline{D}(T)$ and $\overline{\gamma}(T)$ obey the classical
fluctuation-dissipation theorem at low temperatures \cite{kubo}.

\indent
In what follows we shall define a function $S(\omega, \omega')$ which
will, in analogy to the spectral function $J(\omega)$ of the standard
model \cite{caldeira}, allows to replace all the summations
over $k$ by integrals over frequencies:

\begin{equation}
S(\omega, \omega') = \sum^\infty_{n,m=1} g^2_{nm} \delta (\omega -
\Omega_n) \delta (\omega' - \Omega_n)
\end{equation}

\indent
Notice, however, that unlike $J(\omega)$ in \cite{caldeira}, this new
function $S(\omega,\omega')$ is related to the scattering of the
environmental excitations between states of frequencies $\omega$
and $\omega'$ (as seeing from the laboratory frame). Morover, due to
(3.24) it is easy to see that

\begin{equation}
S(\omega, \omega') = S(\omega', \omega)
\end{equation}

\noindent
 From now onwards we shall call $S(\omega,\omega')$ the ``scattering
function".

\indent
Notice that we can rewrite (4.43) and (4.45) as

\newpage

\[
\gamma(t) = \frac{\hbar \theta(t)}{2M_o} \int^\infty_0{d \omega}
\int^\infty_0{d \omega'} S(\omega, \omega') (\omega - \omega')
(n(\omega) - n(\omega'))
\]
\begin{equation}
 cos(\omega - \omega')t
\end{equation}

\noindent
and

\[
D(t) = -\frac{\hbar^2 \theta(t)}{2} \int^\infty_0{d \omega}
\int^\infty_0{d \omega'}
 S(\omega - \omega')(\omega - \omega')^2
 \]
 \begin{equation}
(n(\omega) + n(\omega'))cos(\omega - \omega')t
\end{equation}

\indent
Concluding, we have established that the Hamiltonian (3.26) leads to a
brownian dynamics, that is, the polaron moves as a particle in a
viscous environment where its relaxation and diffusion are due to the
scattering of phonons.

\renewcommand{\theequation}{\thesubsection.\arabic{equation}}
\setcounter{equation}{0}

\section{MOBILITY AND DIFFUSION}

\indent
Equations (4.43) and (4.45) show that the polaron transport
properties depend essentially on the coupling parameter $g_{nm}$.
 From (3.17) we see that this parameter can be obtained if we know the
eigenfunctions of (3.3) and (3.4).

\indent
First of all we can show that (3.3) and (3.4) have solutions with definite
parities. This is easily seen by changing $x$ by $-x$ in (3.3) and (3.4)
and $x'$ by $-x'$ in the integral term.
 From (2.3.32) we observe that $F(-x,-x') = F(x,x')$
and therefore $u_n(x)$ and $u_n(-x)$ obey the same eigenvalue
equation. In other words, the Hamiltonians commute with the parity
operator and therefore it is possible to classify their eigenfunctions
as odd or even. Now we must study the optical and the acoustical cases
separately.

\subsection{Optical Case}

\setcounter{equation}{0}

\indent
Turkevich and Holstein \cite{turke} obtained the exact solutions for
(3.3). For the odd modes the eigenfunctions are:

\begin{equation}
u_n(x) = \sqrt{\frac{ag}{2}} \left(\frac{2n+5}{(n+2)(n+3)}\right)^{1/2}
 (1-Y^2(x)) \frac{dP_{n+2}}{dY}
 \end{equation}
 \[
n = 0,2,4,6...
\]

\noindent
where

\begin{equation}
Y(x) = tanh \left(\frac{gx}{2}\right)
\end{equation}

\noindent
and $P_n$ are the Legendre polynomials.

\indent
The eigenvalues of the problem are

\begin{equation}
\Omega_n = \omega_o \left(1 - \frac{4}{n^2 + 5n + 4}\right)^{1/2}
\end{equation}

\indent
In particular the zero mode, $n = 0$ and $\Omega_o = 0$, is given by (3.5).

\indent
The even modes can be written as

\begin{equation}
u_\alpha(x) = \sqrt{\frac{ag}{2}} \left(\frac{2\alpha +
5}{(\alpha+2)(\alpha+3)}\right)^{1/2} \frac{\left(1-Y^2(x)\right)}{2}
\frac{d}{dY} (P_{\alpha+2}(Y)-P_{\alpha+2}(-Y))
\end{equation}

\noindent
where the allowed values of the $\underline{\alpha}$ are solutions of

\[
\psi(\alpha+3) - \psi(1) = \frac{\pi}{2} tan\left(\frac{\alpha \pi}{2}\right)
\]

\noindent
and $\psi$ is the digamma function. Its eigenvalues are given by

\[
\Omega_\alpha = \omega_o \left(1 - \frac{4}{\alpha^2 + 5 \alpha +
4}\right)^{1/2}
\]

\noindent
We will use the convention given in table 1.

\begin{table}[h]

\begin{tabular}{|c|c|c|c|} \hline
\multicolumn{1}{|c|}{$n$} & \multicolumn{1}{|c|}{$\alpha$} &
\multicolumn{1}{|c|}{$\Omega_n/\omega_o$} &
\multicolumn{1}{|c|}{$\epsilon_n=n-\alpha$} \\ \hline
0 & 0 & 0 & 0  \\
1 & 0.523 & 0.648 & 0.477 \\
2 & 2 & 0.882 & 0 \\
3 & 2.601 & 0.912 & 0.394 \\
4 & 4 & 0.949 & 0 \\
5 & 4.648 & 0.958 & 0.352 \\
6 & 6 & 0.971 & 0 \\
7 & 6.674 & 0.975 & 0.326 \\
8 & 8 & 0.981 & 0 \\
9 & 8.692 & 0.983 & 0.308 \\
10 & 10 & 0.987 & 0 \\ \hline
\end{tabular}

\caption{Conventions for classification of the eigenfunctions}

\end{table}

\indent
As the labels for the even solutions are not integers we define,
$\epsilon_n = n-\alpha$, as the difference between our classification
and the label. Table 1 shows to us that the eigenfrequencies go
quickly to $\omega_o$ while $\epsilon_n$ goes to zero.

\indent
 From (3.17) we note that $G_{nm}$ only couples functions with opposite
parity. Substituting (5.1.1) and (5.1.4) in (3.17) we get:

\begin{equation}
G_{nm} = -\frac{2g}{\pi} \,
\frac{sin(\pi(n-m+\epsilon_m))}{\left[(n-m+\epsilon_m)^2-1\right]} K_{nm}
\end{equation}

\noindent
where

\begin{equation}
K_{nm} =
%% FOLLOWING LINE CANNOT BE BROKEN BEFORE 80 CHAR
\frac{[(n+2)(n+3)(2n+5)(m-\epsilon_m+2)(m-\epsilon_m+3)(2m-2\epsilon_m+5)]^{1/2}
   }{[(n+m+\epsilon_m)^2
+ 10(n+m-\epsilon_m) + 24]}
\end{equation}

\noindent
for: $n = 0,2,4,6 ...$
$m = 1,3,5,7 ...$

\indent
Observe that (5.1.5) is strongly peaked around $n = m \pm 1$. Therefore
the most important contributions to summations involving $G_{nm}$ will
come from these forms (observe that $\epsilon_m$ goes to zero as m goes
to infinite).

\begin{equation}
G_{nm} \simeq g K_{nm} [\delta(n-m-1) - \delta(n-m+1)]
\end{equation}

\noindent
for $n$ even and  $m$ odd.

\indent
 From (4.47) and (3.27) we get:

\[
S(\omega, \omega') = - \frac{g^2}{4} \sum_n \{ C^2_{nn-1}
[\delta(\omega-\Omega_n) \delta(\omega' - \Omega_{n-1}) +
\]
\[
\delta(\omega-\Omega_{n-1}) \delta(\omega'-\Omega_n)] +
+ C^2_{nn+1} [\delta(\omega-\Omega_n) \delta(\omega'-\Omega_{n+1})
\]
\begin{equation}
+ \delta(\omega-\Omega_{n+1}) \delta(\omega'-\Omega_n)]\}
\end{equation}

\noindent
where n is even and

\begin{equation}
C_{nm} = \frac{(\Omega_n + \Omega_m)}{\sqrt{\Omega_n \Omega_m}} K_{nm}
\end{equation}

\noindent
Substituting (5.1.8) in (4.48) we find:

\[
\gamma(t) = -\frac{\hbar g^2}{8M_o} \theta(t)
\sum_n \{ C^2_{nn-1}
(\Omega_n-\Omega_{n-1})(n(\Omega_n)-n(\Omega_{n-1})) cos
(\Omega_n-\Omega_{n-1})t
\]
\begin{equation}
+ C^2_{nn+1}(\Omega_{n+1}-\Omega_n)(n(\Omega_{n+1}-n(\Omega_n))
cos((\Omega_{n+1}-\Omega_n) t)\}
\end{equation}

\indent
We will define wavevectors for each n in (5.1.2) in the form

\begin{equation}
K = \frac{n \pi}{L}
\end{equation}

\noindent
where $L$ is the length of quantization $(L \rightarrow \infty)$.
(5.1.10) then becomes

\begin{equation}
\gamma(t) = - \frac{\hbar^2g^2}{8M_o} \theta(t) \frac{2 \pi}{L}
\int^\infty_0{dK} C^2(K,K) \left(\frac{d \Omega}{dK}\right)^2
\left( \frac{dn}{d\Omega}\right)
 cos\left(\frac{\pi}{L} \frac{d \Omega}{dK} t\right)
\end{equation}

\noindent
where we used the limit $L \rightarrow \infty$.

\indent
It is easy to see that by (5.1.3)

\begin{equation}
\frac{d \Omega}{dK} = \frac{4\pi^2}{L^2} \frac{\omega_o}{K^3} \; as \, L
\rightarrow \infty
\end{equation}

\noindent
and by (5.1.9) and (5.1.6) that

\[
C^2(K,K) = \frac{L^2K^2}{\pi^2} \; as \, L \rightarrow \infty .
\]

\indent
Now, using the fact that the frequencies approach very fast the value
$\omega_0$ when $n$ increases we make the following approximation:

\[
\frac{dn}{d \Omega} \simeq \frac{dn}{d \Omega} \mid_{\Omega =
\omega_o}
\]

\noindent
we shall rewrite (5.1.12) as

\[
\gamma(t) = \frac{\hbar^2 g^2}{4M_o} \theta(t) \left(- \frac{dn}{d \Omega}
\mid_{\omega_o}\right)
\omega^2_o \frac{16 \pi^3}{L^3} \int^\infty_0{dK} \frac{cos
\left(\frac{4 \pi^3 \omega_ot}{L^3} \frac{1}{K^3}\right)}{K^4}
\]

\indent
This integral can be easily done if we change variables, $x = \frac{4
\pi^3 \omega_o}{L^3} \frac{1}{K^3}$. The integral then becomes

\[
\gamma(t) = \frac{\hbar^2 g^2}{4M_o} \theta(t) \left(-\frac{dn}{d \Omega}
\mid _{\omega_o}\right) \frac{4 \omega_o}{3} \int^\infty_0{dx} \; cos (xt)
\]

\noindent
and finally

\begin{equation}
\gamma(t) = \frac{\hbar g^2}{2M_o} \frac{\pi}{3} \frac{\hbar
\omega_o}{K_B T} \,
e^{\hbar \omega_o/K_B T}\,\left(e^{\hbar \omega_o/K_B T} - 1\right)^{-2}
\delta(t)
\end{equation}

\noindent
which has the form (4.44).

\indent
For low temperatures, $K_BT \ll \hbar \omega_o$, we have

\begin{equation}
\overline{\gamma}(T) \simeq \frac{\pi}{6} \frac{\hbar g^2}{M_o}
\left(\frac{\hbar \omega_o}{K_B T}\right) e^{-\hbar \omega_o/K_B T}
\end{equation}

\noindent
so $\overline{\gamma}(T)$ goes to zero as T goes to zero, as expected.
So, for very low temperatures the mobility is extremely high and the polaron
moves as a free particle. This is an expected result since at $T=0$ there
are no phonons to be scattered.

 \indent
For high temperatures, $K_BT \gg \hbar \omega_o$,

\begin{equation}
\overline{\gamma}(T) \simeq \frac{\pi}{6} \frac{\hbar g^2}{M_o}
\left(\frac{K_B T}{\hbar \omega_o}\right)
\end{equation}

\noindent
and the mobility goes to zero as $T \rightarrow \infty$.

\indent
The diffusion parameter (4.49) can be calculated in the same way. It gives

\begin{equation}
\overline{D}(T) = \frac{\pi}{3} \frac{\hbar g^2}{M_o} \hbar \omega_o
\left(e^{\hbar \omega_o/K_BT} - 1\right)
\end{equation}

\indent
For low temperatures this parameter is too small, going to zero as $T
\rightarrow 0$. The fluctuations are once again small due to the absence of
phonons. For high temperature we see that the fluctuations also increase
linearly with $T$.

\indent
Observe that

\begin{equation}
\frac{\overline{D}(T)}{\overline{\gamma}(T)} = 2 M_o K_B T \left(1 -
e^{-\hbar \omega_o /K_B T}\right)
\end{equation}

\noindent
which gives the classical result of the fluctuation-dissipation
theorem \cite{kubo} for the brownian motion at low temperatures.

\subsection{The Acoustical Case}

\setcounter{equation}{0}

\indent
As we do not have exact solutions for (3.4) it will be necessary to
make some approximations in the present analysis. Observe that (3.4)
is a Schr\"{o}dinger-like equation for a particle in a non-local potential

\[
V(x,x') = -\frac{\partial^2 F(x,x')}{\partial x' \partial x}
\]

\indent
 From (2.3.32) we observe that $V(x,x')$ goes to zero as $x$ goes to
infinity. Actually, the potential is almost zero except in the range

\[
-\frac{1}{g} < x < \frac{1}{g}
\]

\indent
Out of this range the wave function can be well described by

\begin{equation}
\frac{d^2 u_k}{dx^2} + k^2 u_k(x) = 0
\end{equation}

\noindent
where we have used that

\begin{equation}
\omega = v_s \mid k \mid
\end{equation}

\noindent
The solutions of (5.2.1) must be classified as even or odd. We choose

\begin{equation}
u_E(k,x) = \sqrt{\frac{2a}{L}} cos(k \mid x \mid + \delta_E(k))
\end{equation}
\begin{equation}
u_O(k,x) = \sqrt{\frac{2a}{L}} \, sgn(x) \, sin(k \mid x \mid + \delta_O(k))
\end{equation}

\noindent
where

\[
sgn(x) = \left\{ \begin{array}{l}
1 \, \,if \, \,x > 0 \\
-1\, \,if \, \,x < 0
\end{array} \right.
\]

\noindent
and $\delta_E(k)$ and $\delta_O(k)$ are the phase shifts for the even
and odd modes, respectively, which must appear due to the presence of
the potential.

\indent
Another possible solution of (5.2.1) is:

\begin{equation}
u_k(x) = \sqrt{\frac{2a}{L}} \; \left\{t(k) e^{ikx} \theta(x-1/g)
+ \left(e^{ikx} + r(k) e^{-ikx}\right) \theta(-x - 1/g)\right\}
\end{equation}

\noindent
This expression can be interpreted as a wave incident from the left on
a potential whose $t(k)$ and $r(k)$ are the transmition
and reflection amplitudes.

\indent
We can construct (5.2.5) from (5.2.3) and (5.2.4) as \cite{lipkin}

\[
u_k(x) = e^{i \delta_E}u_E(k,x) + i e^{i\delta_O} u_O(k,x)
\]

\noindent
if

\[
t(k) = \frac{1}{2} \left(e^{2i \delta_E(k)} + e^{2i \delta_O(k)}\right)
\]
\[
r(k) = \frac{1}{2} \left(e^{2i \delta_E(k)} - e^{2i \delta_O(k)}\right)
\]

\noindent
Consequently the transmission and reflection coefficients are given by

\begin{equation}
T(k) = \mid t(k) \mid^2 = cos^2 (\delta_O(k) - \delta_E(k))
\end{equation}
\begin{equation}
R(k) = \mid r(k) \mid^2 = sin^2 (\delta_O(k) - \delta_E(k))
\end{equation}

\noindent
and $T(k) + R(k) = 1$ as expected.

\indent
Once we have the phase shifts of the problem we can find the
transmission and reflection probabilities using (5.2.6) and (5.2.7), or
alternatively, if we have the reflection and transmission amplitudes we
can obtain the phase shifts

\begin{equation}
\delta_E(k) = \frac{1}{2} arctan \left[\frac{Im(t(k) + r(k))}{Re(t(k)
+ r(k))}\right]
\end{equation}
\begin{equation}
\delta_O(k) = \frac{1}{2} arctan \left[\frac{Im(t(k) - r(k))}{Re(t(k) - r(k))}
\right]
\end{equation}

\indent
Actually, Sch\"{u}ttler and Holstein \cite{schu} obtained these
coefficients in the limit of long and short wavelength after a rather
intricate algebra: \footnote{ In the results of ref. 5, R and T depend
on the polaron velocity which is very small in our case and we have
put it equal to zero.}

\noindent
i) for $k \gg g$

\begin{equation}
r(k) \simeq \frac{16 \pi^2 i k^3 e^{-2\pi k/g}}{g^3}
\end{equation}
\begin{equation}
t(k) \simeq 1 + \frac{2 i g}{5k}
\end{equation}

\noindent
ii) for $k \ll g$

\begin{equation}
r(k) \simeq - \frac{3 i k}{g} - \frac{gk^2}{g^2}
\end{equation}
\begin{equation}
t(k) \simeq 1 - \frac{3 i k}{g} - \frac{g k^2}{g^2}
\end{equation}

\indent
These results allow one to compute the respective phase shifts as:

\noindent
i) for $k \gg g$

\begin{equation}
\delta_E(k) \simeq \delta_O(k) \simeq \frac{g}{5k}
\end{equation}

\noindent
ii) for $k \ll g$

\[
\delta_E(k) \simeq -\frac{3k}{g}
\]
\begin{equation}
 \delta_O(k) \simeq 0
\end{equation}

\indent
So, the phase shifts are very small. We would say that there is a
propagation of sound waves through the polaron.

\indent
If the interaction between the electron and the lattice is strong,
the range of the potential is small. Therefore, the contribution to the
integral in (3.17) due to the true solution is almost the same as the one
we would have got had we used the free solutions (5.2.3) and (5.2.4).

\indent
Firstly, we impose periodic boundary conditions which gives the
allowed values for $k$:

\begin{equation}
k_n = \frac{2 n \pi}{L} \;\; n = \pm 1, \pm 2, \pm 3 \ldots
\end{equation}

\indent
In order to classify the solutions we will use the following convention:

\begin{equation}
u_{2n-1}(x) = u_E(n,x)
\end{equation}
\begin{equation}
u_{2n}(x) = u_O(n,x)
\end{equation}

\noindent
for $n = \pm 1, \pm 2, \pm 3 \ldots$

\indent
Now we can evaluate (3.17). It yields

\[
G_{2n-1,2m} = -\frac{2 k_n}{L} \{ \frac{sin[(k_n - k_m)L/2]}{k_n - k_m}
cos (\delta_E(k_n) - \delta_O(k_m))
\]
\begin{equation}
+ \left[\frac{1 - cos[(k_n-k_m)L/2]}{k_n-k_m} \right]
 sin(\delta_E(k_n)-\delta_O(k_m)) \}
\end{equation}

\indent
As in the optical case we have a matrix with zeros in the diagonal
and with off-diagonal terms which decrease as a function of their
distance to the main diagonal.

\indent
When $L \rightarrow \infty$ we will have (using (5.2.6))

\begin{equation}
G_{kk'} = -\frac{2k}{L} \left\{\pi \delta (k-k') \sqrt{T(k)} +
P\left[\frac{sin(\delta_E(k) - \delta_O(k'))}{k-k'}\right] \right\}
\end{equation}

\noindent
where $P$ denotes the principal value.

\indent
Substituting (5.2.20) in (4.47), transforming the summations into integrals
and using (5.2.2) we get

\begin{equation}
S(\omega, \omega ') = -\frac{2L}{v^2_s} \omega^2 \sqrt{T(\omega)}
\delta(\omega - \omega ')
- \frac{1}{4\pi^2 v^2_s} \frac{\omega(\omega+\omega ')}{\omega
'(\omega-\omega')^2} sin^2 (\delta_E(\omega)-\delta_O(\omega '))
\end{equation}

\indent
In (4.48) we will change the variables of integration and rewrite (4.47) as

\[
\gamma(t) = \frac{\hbar \theta(t)}{2M_O} \int^{\omega_D}_0{d \theta}
\int^{\omega_D}_{-\omega_D}{d \Omega} \;  S\left(\theta + \frac{\Omega}{2},
\theta-\frac{\Omega}{2}\right) \;  \Omega
\]
\begin{equation}
\left(n\left(\theta+\frac{\Omega}{2}\right) -
 n\left(\theta-\frac{\Omega}{2}\right)\right) \, cos(\Omega t)
\end{equation}

\noindent
where

\[
\Omega = \omega - \omega'
\]
\[
\theta = \frac{\omega + \omega '}{2}
\]

\indent
Observe that we have replaced the limit on the integration by the cut-off
frequency, $\omega_D$.

\indent
Actually we are interested in a time scale, $\tau$, which is much
longer than the typical phonon period or

\[
\tau \gg \omega^{-1}_D
\]

\indent
With this approximation the cosine term in (5.2.22) oscillates rapidly,
giving no contribution to the integration, except when $\Omega$ is
close to zero. So we can approximate (5.2.22) as

\begin{equation}
\gamma(t) = \frac{\hbar \pi \delta(t)}{2m} \int^\infty_0{d \theta}
f(\theta) \left(- \frac{dn}{d \theta}\right)
\end{equation}

\noindent
where

\begin{equation}
f(\theta) = -\lim_{\epsilon \rightarrow 0} \epsilon^2 S\left(\theta +
\frac{\epsilon}{2}, \theta - \frac{\epsilon}{2}\right)
\end{equation}

\noindent
Now, using (5.2.21) and noticing that the delta term does not
contribute to (4.48) we get

\begin{equation}
f(\theta) = \frac{1}{4 \pi^2 v^2_s} \theta^2 R(\theta)
\end{equation}

\noindent
where we have used (5.2.7).

\indent
So, we conclude that we have here a Markovian process with the damping
parameter given by

\begin{equation}
\overline{\gamma}(T) = \frac{\hbar^2}{8 \pi M_O v^2_s K_B T}
\int^{\infty}_0{d \omega} \; \omega^2 \; R(\omega) \; \frac{e^{\hbar \omega/K_B
T}}{\left(e^{\hbar \omega/K_BT}-1\right)^2}
\end{equation}

\indent
Defining a new variable

\[
\omega = g v_s \kappa/2
\]

\noindent
and a typical phonon temperature, $T_c$, by

\[
T_c = \frac{\hbar g v_s}{2 K_B}
\]

\noindent
we can evaluate (5.2.26) which reads

\begin{equation}
\overline{\gamma}(T) = \frac{\hbar g^2}{32 \pi M_O}
I\left(\frac{T_c}{T}\right)
\end{equation}

\noindent
where

\begin{equation}
I(S) = S \int^{\infty}_0{d\kappa} \; \kappa^2 \; R(\kappa) \;
\frac{e^{S\kappa}}
   {(e^{S\kappa}-1)^2}
\end{equation}

\noindent
is exactly the result obtained by Sch\"{u}ttler and Holstein
\cite{schu} for the polaron mobility using the kinetic theory.

\indent
For small temperatures, $T \ll T_c$, we shall use the long wavelength
reflectivity (see (5.2.12)) since it gives the largest contribution
to the occupation number

\begin{equation}
R(\kappa) \simeq \frac{9}{4} \kappa^2
\end{equation}

\noindent
Then, using (5.2.28) and (5.2.29) we can approximate (5.2.25) by

\begin{equation}
\overline{\gamma}(T) = \frac{27 \hbar g^2}{16 \pi M_O} \left(
 \frac{T}{T_c} \right)^4
\end{equation}

\indent
This result shows that the acoustical polaron, as the optical one,
behaves as a free particle as $T \rightarrow 0$.

\indent
For high temperatures, $T \gg T_c$, we use the expression for short
wavelengths (see (5.2.10))

\begin{equation}
R(\kappa) \simeq 4 \pi^4 \kappa^6 e^{-2\pi \kappa}
\end{equation}

\noindent
and one has

\begin{equation}
\overline{\gamma}(T) = \frac{315 \hbar g^2}{64 \pi^4 M_O}
\left( \frac{T}{T_c} \right) ,
\end{equation}

\noindent
which means that the mobility decreases for high temperatures.

\indent
We can calculate the diffusion coefficient (4.49) in the same way and
we get

\begin{equation}
\overline{D}(T) = \frac{\hbar g^2}{32 \pi M_O} K_B T_c J(S)
\end{equation}

\noindent
where

\begin{equation}
J(S) = \int^\infty_0{d\kappa} \, \frac{\kappa^2 \; R(\kappa)}{e^{S\kappa} - 1}
\end{equation}

\noindent
and we have used the fact that the diffusion is a Markovian process.

\indent
For small temperatures, $T \ll T_c$, the diffusion coefficient is
given by:

\begin{equation}
\overline{D}(T) \simeq \frac{27 g^2 \hbar}{16 \pi} K_B \frac{T^5}{T^4_c}
\end{equation}

\noindent
and the fluctuations decrease very fast as the temperature is lowered,
exactly as in the optical case.

\indent
So, the relation between relaxation and diffusion is the classical one
for the brownian motion \cite{kubo}:

\begin{equation}
\frac{\overline{D}(T)}{\overline{\gamma}(T)} = M_O K_B T
\end{equation}

\indent
For high temperature, $T \gg T_c$,

\begin{equation}
\overline{D}(T) = \frac{315 \hbar g^2}{16 \pi^5} K_B T
\end{equation}

\indent
And exactly as in the optical case the fluctuations increase linearly
with temperature, this is the classical result \cite{kubo} which is
expected to be valid in the high temperature limit.

\section{CONCLUSIONS}

\indent
In the foregoing sections we have shown that the semiclassical (mean
field) method enables us to visualize the polaron physics and allows
us to treat the strong coupling limit of an electron interacting with
a lattice.

\indent
The advantage of dealing with this method is the fact that, in terms
of the coordinate and the modified phonons, we reduce the problem to
a new model for treating quantum dissipation. In a sense, the non-linear
character of the electron-phonon interaction is somehow ``hidden" in
the soliton-like solution whose center is regarded as the polaron
coordinate.

\indent
Eliminating the electron operators by perturbative techniques
(that is, tracing over the electron coordinates ) and
using the well known collective coordinate formalism we get an
effective Hamiltonian for the polaron in the presence of renormalized
phonons. That Hamiltonian, in the approximation of small polaron
velocity, is reduced to a very simple form which takes into account only
processes which involve polaron-phonon collisions.

\indent
We developed a new functional method to treat the Hamiltonian in the
limit of small polaron's velocity. Our method showed that the polaron
moves as a brownian particle which collides with the light particles
of the environment. This method provided us with a tool for a systematic
calculation
of the damping parameter (and, as a consequence, the mobility) and the
diffusion coefficient as function of the temperature. We have also showed
that in the time scale of interest the motion is essentially
Markovian, that is, it does not have memory.

\indent
An important comment about our work is that it is fully quantized
and the ``semiclassical" argument is only used as an artifact.
Furthermore, it confirms some important results for the acoustical
polaron obtained by Sch\"{u}ttler and Holstein \cite{schu} using
kinetic transport theory.

\indent
We are very grateful to Professor A.J. Leggett for a critical reading of
the manuscript.
One of us (A.H.C.N.) would like to acknowledge FAPESP (Fundacao de Amparo
a Pesquisa
do Estado de Sao Paulo) for a scholarship, while (A.O.C.) kindly acknowledges
the partial support of CNPq (Conselho Nacional de Desenvolvimento Cientifico
e Tecnologico) and FAEP (Fundo de Apoio ao Ensino e Pesquisa da UNICAMP).

\newpage

\renewcommand{\theequation}{\arabic{equation}}
\setcounter{equation}{0}

\section*{APPENDIX A}

\indent
In this appendix we wish to calculate the first order correction in
energy due to a many electron wave function in (2.3.12) and (2.3.17).

\indent
In case of many electrons the ground state is the Fermi sphere with
radius $k_F$, which is given by

\begin{equation}
k_F = \frac{\pi}{2a} \left(\frac{N_e}{N}\right)
\end{equation}

\noindent
where $N_e$ is the number of electrons and $N$ the number of sites.

\indent
The non-perturbed Hamiltonian is

\begin{equation}
H_O = E_O a^{\dag} a + \sum_k E_n a^{\dag}_n a_n
\end{equation}

\noindent
with the ground state wave function

\begin{equation}
\mid \psi^o > = \mid n_k = 1, k \leq k_F ; n_k = 0, k > k_F \, > \, \, .
\end{equation}

\noindent
The ground state energy is

\begin{equation}
E^o = \frac{N}{\hbar} \frac{\hbar^2}{ma^2} (k_F a)^3 - \frac{\hbar^2
g^2}{4m}
\end{equation}

\noindent
where we have accounted for the spin degeneracy.

\indent
The first order correction is given from the interaction term in (2.3.12)
or (2.3.17)

\[
E^{(1)} = < \psi_O \mid H_I \mid \psi_O >
\]

\noindent
For the optical case it reads

\[
E^{(1)} = \frac{D}{a} \sum_{qk} f_{qq}(k) \theta(k_F - \mid q \mid) q_K
\]

\noindent
or

\begin{equation}
E^{(1)} = \sum_k \frac{\Delta (k)}{a} q_k
\end{equation}

\noindent
where

\begin{equation}
\Delta(k) = \frac{D}{\pi \sqrt{N}} \left\{k_F \delta(k) - \frac{2 \pi}{g}
 arctan\left(\frac{2 k_F}{g}\right) \mid k \mid csch\left(\frac{2 \mid k
\mid}{g}\right)\right\}
\end{equation}

\noindent
For the acoustical case,

\begin{equation}
E^{(1)} = i \sum_k k \Delta(k) q_k
\end{equation}

\indent
So, at first order the ions are displaced from their equilibrium
positions by

\[
\frac{\Delta(-k)}{Ma \omega^2_O}
\]

\noindent
in the optical case and

\[
\frac{i k \Delta(-k)}{M \omega^2_k}
\]

\noindent
in the acoustical case. This causes a change in the energy given by

\[
\Delta E = - \sum_k \frac{\Delta(k) \Delta(-k)}{2 a^2 M \omega^2_O}
\]

\noindent
for the optical case, and

\[
\Delta E = - \sum_k \frac{\Delta(k) \Delta(-k)}{2 M v^2_s}
\]

\noindent
for the acoustical case.

\setcounter{equation}{0}

\section*{APPENDIX B}

\indent
Let us first show how to obtain the zero mode solution for the optical
case. Making $\Omega_0 = 0$ in (3.3) we get

\begin{equation}
u_O(x) = \int{dx'} F(x,x') u_O(x')
\end{equation}

\noindent
Define a function $g(x,x')$

\begin{equation}g(x,x') = \sum^\infty_{n=0} \frac{\phi^{*F}_n (x)
\phi^F_n (x')}{\left(k^2_n + g^2/4\right)}
\end{equation}

\noindent
 Then, from (2.3.1)

\[
\left(- \frac{\partial^2}{\partial x^2} - \frac{g^2}{2} sech^2
\left(\frac{gx}{2}\right) + \frac{g^2}{4}\right) g(x,x') =
\]
\begin{equation}
 = - \frac{g}{4} sech\left(\frac{gx}{2}\right) sech\left(\frac{gx'}{2}\right)
 +\delta(x,x')
\end{equation}

\noindent
where we have used the completeness of the adiabatic states

\begin{equation}
\phi_O(x) \phi_O(x') + \sum^\infty_{n=0} \phi^{F^*}_n (x)
\phi^F_n(x') = \delta(x,x')
\end{equation}

\noindent
and the explicit form for $\phi_O(x)$.

\indent
Let us rewrite (1) as

\begin{equation}
u_O(x) = g^2 sech\left(\frac{gx}{2}\right) f(x)
\end{equation}

\noindent
where

\begin{equation}
f(x) = \int{dx'} g(x,x') sech\left(\frac{gx'}{2}\right) u_O(x')
\end{equation}

\noindent
Then, from (3)

\[
\left[\frac{d^2}{dx^2} + \frac{g^2}{4}
 \left(2 \, sech^2 \left(\frac{gx}{2}\right)
 - 1\right)\right] f(x)
\]
\begin{equation}
= sech\left(\frac{gx}{2}\right) \left[- u_O(x) + \frac{g}{4}
 \int{dx'} sech^2\left(\frac{gx'}{2}\right) u_O(x')\right]
\end{equation}

\noindent
The last term in (7) must vanish because $u_O(x)$ is odd. Using
(5) we get

\begin{equation}
\left[\frac{d^2}{dx^2} + \frac{g^2}{4} \left(
 6 \, sech^2\left(\frac{gx}{2}\right)
 - 1\right)\right] f(x) = 0
\end{equation}

\noindent
The solution is easily obtained \cite{morse} and reads

\begin{equation}
f(x) = sech \left(\frac{gx}{2}\right) tanh \left(\frac{gx}{2}\right)
\end{equation}

\noindent
Substituting (9) in (5) and normalizing it, we get

\begin{equation}
u_O(x) = \sqrt{\frac{15 ag}{2}} tanh \left(\frac{gx}{2}\right) sech^2
\left(\frac{gx}{2}\right)
\end{equation}

\indent
For the acoustical case we must use (3.4) with $\Omega_0 = 0$, which
reads

\begin{equation}
\frac{d^2 u_O}{dx^2} = \frac{d}{dx} \int{dx'} F(x,x')
\frac{du_O}{dx'}(x')
\end{equation}

\indent
Now, using (2) we should define

\begin{equation}
h(x) = \int{dx'} g(x,x') sech \left(\frac{gx'}{2}\right) \frac{du_O}{dx'} (x')
\end{equation}

\indent
and rewrite (11) as

\[
\frac{d^2 u_O}{dx^2} = g^2 \frac{d}{dx} \left\{sech (\frac{gx}{2})
 h(x)\right\}
\]

\indent
This can be easily integrated yielding

\begin{equation}
u_O(x) = g^2 \int^x_{-\infty}{dx'} sech \left(\frac{gx'}{2}\right)
 h(x')
\end{equation}

\noindent
where

\[
u_O(-\infty) = 0
\]

\noindent
Observe that $h(x)$ also obey eq. (8), so

\begin{equation}
h(x) = tanh \left(\frac{gx}{2}\right) sech \left(\frac{gx}{2}\right)
\end{equation}

\noindent
Substituting (14) in (13) we get, after normalization

\begin{equation}
u_O(x) = \sqrt{\frac{3 ag}{8}} sech^2 \left(\frac{gx}{2}\right)
\end{equation}

\setcounter{equation}{0}

\section*{APPENDIX C}

\indent
We shall evaluate here the functional form for the superpropagator, $J$.
 From (3.26) we see that the Hamiltonian can be put in the form

\begin{equation}
H = H_S + H_R + H_I
\end{equation}

\noindent
where

\begin{equation}
H_S = P^2/2M_O
\end{equation}
\begin{equation}
H_R = \sum^\infty_{n=1} \hbar \Omega_n b^{\dag}_n b_n +
(\sum^\infty_{n,m=1} \hbar g_{nm} b^{\dag}_m b_n)^2/2 M_O
\end{equation}
\begin{equation}
H_I = -P \sum^\infty_{n,m=1} \hbar g_{nm} b^{\dag}_m b_n/M_O
\end{equation}

\noindent
 From (4.6) we get

\[
<x \mid \hat{\rho}_s (t) \mid y > = \int \cdots
\int{\left(\prod^N_{k=1} \frac{d^2 \alpha_k}{\pi^N}\right)}
 <x \vec{\alpha} \mid
\hat{\rho}(t) \mid y \vec{\alpha} >
\]

\noindent
or using the completeness relations (4.10) and (4.11) we can, with the
help of (4.2), write

\begin{equation}
\rho_s(x,y,t) = \int{dx'} \int{dy'} \, \rho_s(x',y',0) J(x,y,t;x',y',0)
\end{equation}

\noindent
where

\begin{equation}
J = \int{\frac{d^2 \vec{\alpha}}{\pi^N}} \int{\frac{d^2
\vec{\beta}}{\pi^N}} \int{\frac{d^2 \vec{\beta'}}{\pi^N}}
\rho_R(\vec{\beta}^*, \vec{\beta}',0)
K(x \vec{\alpha}^* x' \vec{\beta}, t) K^*(y \vec{\alpha} y'
\vec{\beta}^{'*} t)
\end{equation}

\noindent
with

\begin{equation}
K(x \vec{\alpha}^* x' \vec{\beta} t) = <x \vec{\alpha} \mid e^{-i
\hat{H} t/\hbar} \mid x' \vec{\beta} >
\end{equation}

\noindent
In order to transform (7) into a functional integral we must divide t in
(M-1) subintervals of length $\epsilon$ and use (M-1) completeness relations
between the (M-1) exponentials in (7). Then

\[
< x \vec{\alpha} \mid e^{-i \hat{H} t/\hbar} \mid x' \vec{\beta} > =
 = \int{dq_{N-1}} \ldots \int{dq_1} \int{\frac{d^2
\vec{\alpha}_{N-1}}{\pi^N}} \ldots
\int{\frac{d^2 \vec{\alpha}_1}{\pi^N}}
\]
\begin{equation}
 <q_M \vec{\alpha}_M \mid
e^{-i \hat{H} \epsilon/\hbar} \mid q_{M-1} \vec{\alpha}_{M-1} >
\cdots
 <q_1 \vec{\alpha}_1 \mid e^{-i \hat{H} \epsilon/\hbar} \mid q_0
\vec{\beta}_O >
\end{equation}

\noindent
where

\begin{equation}
q_M = x , \vec{\alpha}^*_M = \vec{\alpha}^* , q_0 = x',
\vec{\alpha}_O = \vec{\beta}  \, \, .
\end{equation}

\indent
Now insert $M$ completeness relations in the momentum representation

\[
\int{dp} \mid p > < p \mid = 1
\]

\noindent
in (8) in order to obtain

\[
K = \prod^{M-1}_{k=1} \left\{\int{dq_k \frac{d^2 \vec{\alpha}_k}{\pi}} \right\}
\prod^M_{k=1} \left\{\int{dP_k}\right\}
\]
\[
<q_N \vec{\alpha}_N \mid e^{-iH \epsilon/\hbar} \mid P_N
\vec{\alpha}_{N-1} > <P_N \mid q_{N-1} >
\]
\begin{equation}
 \cdots <q_1 \vec{\alpha}_1 \mid e^{-iH \epsilon/\hbar} \mid P_1
\vec{\alpha}_O > < P_1 \mid q_0 >
\end{equation}

\noindent
Now we will take the limit that $M \rightarrow \infty , \epsilon
\rightarrow 0$ but with $t = (M-1) \epsilon$ being finite.

\indent
For small $\epsilon$ we should expand the exponential in (10) to
first order in $\epsilon$ and write

\[
< q_k \vec{\alpha}_k \mid e^{-iH \epsilon/\hbar} \mid P_k
\vec{\alpha}_{k-1} >
 \simeq <q_k \mid P_k > <\vec{\alpha}_k \mid \vec{\alpha}_{k-1} >
\]
\begin{equation}
 exp\left\{- \frac{i}{\hbar} \epsilon H(q_k, P_k, \vec{\alpha}^*_k,
\vec{\alpha}_{k-1})\right\}
\end{equation}

\noindent
where

\begin{equation}
H(q_k P_k \vec{\alpha}^*_k \vec{\alpha}_{k-1}) = \frac{<q_k
\vec{\alpha}_k \mid \hat{H} \mid P_k
\vec{\alpha}_{k-1}>}{<\vec{\alpha} \mid \vec{\alpha}_{k-1} >}
\end{equation}

\indent
Using the overlaping relations

\[
< \vec{\alpha} \mid \vec{\beta} > = exp \left\{ \vec{\alpha}^* \cdot
\vec{\beta} - \frac{\mid \vec{\alpha} \mid ^2}{2} - \frac{\mid
\vec{\beta} \mid ^2}{2} \right\}
\]
\[
<P \mid q > = \frac{1}{2 \pi \hbar} exp \left\{- \frac{i}{\hbar} Pq
\right\}
\]

\noindent
one obtains

\[
K = \int{\frac{dP_M}{2\pi \hbar}} \prod^{M-1}_{k=1} \left\{
\int{\frac{dq_k dP_k}{2\pi \hbar} \frac{d^2 \vec{\alpha}_k}{\pi^N}}
\right\}
\]
\[
exp \{ \sum^M_{k=1} \frac{1}{2} \left[\vec{\alpha}_{k-1}
 \left(\vec{\alpha}^*_k
- \vec{\alpha}^*_{k-1}\right) - \vec{\alpha}^*_k \left(\vec{\alpha}_k -
\vec{\alpha}_{k-1} \right)\right]
\]
\begin{equation}
 + \frac{i}{\hbar} \left[P_k(q_k - q_{k-1}) - \epsilon H(P_k q_k
\vec{\alpha}^*_k \vec{\alpha}_{k-1} )\right] \}
\end{equation}

\noindent
Now one has to integrate over $P_k$. So, using (1) we must evaluate

\begin{equation}
\int{\frac{dP_k}{2\pi \hbar}} \, exp\left\{- \frac{i \epsilon}{\hbar}
\left[\frac{P^2_k}{2M_O} - P_k \left(h + \frac{(q_k-q_{k-1})}{\epsilon}
 \right)\right]\right\}
\end{equation}

\noindent
where

\begin{equation}
h = \sum^\infty_{n,m=1} \frac{\hbar}{M_O} g_{nm} \alpha^*_n \alpha_m
\label{(15)}
\end{equation}

\noindent
This allows one to rewrite (14) in the standart form \cite{feyn}

\begin{equation}
\left(\frac{M_O}{2\pi i \hbar \epsilon}\right)^{1/2}
 exp \left\{\frac{iM_O
\epsilon}{2 \hbar} \left(h + \frac{(q_k - q_{k-1})}{\epsilon}\right)^2 \right\}
\end{equation}

\noindent
Now, substituting (16) in (13) and taking the limit of $M \rightarrow
\infty$ and $\epsilon \rightarrow 0$ we get \cite{feynman}

\begin{equation}
K = \int^{x}_{x'}{Dq}
\int^{\vec{\alpha}^*}_{\vec{\beta}}{D^2\vec{\alpha}} \, e^{-\frac{\mid
\vec{\alpha} \mid^2}{2} - \frac{\mid \vec{\beta} \mid^2}{2}} \, exp\{ S[q,
\vec{\alpha}]\}
\end{equation}

\noindent
where

\begin{equation}
Dq = \lim_{M \rightarrow \infty , \epsilon \rightarrow 0}
\left\{ \left(\frac{M_O}{2\pi i \hbar \epsilon}\right)^{M/2}
 \prod^{M-1}_{k=1} dq_k \right\}
\end{equation}
\begin{equation}
D^2\vec{\alpha} = \lim_{M \rightarrow \infty} \prod^{M-1}_{k=1}
\left(\frac{d^2 \vec{\alpha}_k}{\pi^N} \right)
\end{equation}

\noindent
with

\[
S = \int^t_0{dt'} \left\{\frac{iM_O}{\hbar 2} \left(\dot{q}
 + h(\vec{\alpha}^*,
\vec{\alpha})\right)^2 + \frac{1}{2} \left(\vec{\alpha} \vec{\dot{\alpha}}^* -
\vec{\alpha}^* \vec{\dot{\alpha}} \right)
 - \frac{i}{\hbar} H_R (\vec{\alpha}^*, \vec{\alpha}) \right\}
\]

\noindent
Finally, using (3) and (15) one reaches

\[
S = \int^t_0{dt'} \left\{ \frac{1}{2} \left( \vec{\alpha}
 \vec{\dot{\alpha}}^* -
\vec{\alpha}^* \vec{\dot{\alpha}}\right) + \frac{i}{\hbar}
\left(\frac{M_O \dot{q}^2}{2} + M_O \dot{q} h\right)
- i \sum^\infty_{n,m=1} \Omega_n \alpha^*_n \alpha_n \right\}
\]

\noindent
Now, substituting (21) in (17) and (6) we get the result (4.12).

\end{document}